\documentclass[11pt,a4j]{article}
\usepackage[dvips]{graphicx}
\usepackage{amsmath}
\usepackage{amssymb}
\usepackage{amsfonts}
\usepackage{bm}
\usepackage{enumerate}

\newcommand*{\Lcdot}{\raisebox{-0.25ex}{\scalebox{1.5}{$\cdot$}}}
\begin{document}
\baselineskip=4.75
mm
\centerline{\bf  Hamiltonian structure for  two-dimensional extended Green-Naghdi equations}\par
\bigskip
\centerline{\rm Yoshimasa Matsuno\footnote{Author for correspondence (matsuno@yamaguchi-u.ac.jp).}}\par
\centerline{\it Division of Applied Mathematical Science, Graduate School of Science and Engineering} \par
\centerline{\it Yamaguchi University, Ube, Yamaguchi 755-8611, Japan} \par
\bigskip
\bigskip
\bigskip
\noindent The two-dimensional  Green-Naghdi (GN) shallow-water model   for surface gravity waves is extended to 
incorporate the arbitrary higher-order dispersive effects.  This can be achieved
 by developing a novel asymptotic analysis applied to the basic nonlinear water wave problem.
The linear dispersion relation for the extended GN system is then explored in detail.
In particular, we use its characteristics to discuss the well-posedness of the linearized problem.
As illustrative examples of approximate model equations,  we derive a higher-order model that is accurate to the fourth power of the dispersion parameter in the case of a flat
bottom topography, and address the related issues such as the linear dispersion relation, conservation laws and the pressure distribution at the fluid bottom
on the basis of this model. The original GN model is then briefly described in the case of an uneven bottom topography.
Subsequently, the extended GN system  presented here is shown to have the same Hamiltonian structure as that of the original
GN system. 
Last, we demonstrate that Zakharov's Hamiltonian formulation of surface gravity waves 
is equivalent to that of the extended GN system by rewriting the former system
in terms of the momentum density instead of the velocity potential at the free surface.
  \par

\newpage
\leftline{\bf  1. Introduction} \par
\medskip
\noindent In a recent paper (Matsuno [1]), we  extended the Green-Naghdi (GN) shallow-water model equations to incorporate the arbitrary higher-order dispersive effects while preserving the full nonlinearity.
The system of equations thus obtained is the generalization of the model equations derived first by Serre [2] and later by Su \& Gardner [3] to describe the one-dimensional (1D) propagation of fully nonlinear and 
weakly dispersive surface gravity waves. We showed that the extended GN system has the same Hamiltonian structure as that of the original GN system, 
and it is equivalent to Zakharov's Hamiltonian formulation of surface gravity waves.
Green \& Naghdi [4] were the first to generalize Serre's system to the two-dimensional (2D) case while taking into account the effect of an  uneven bottom topography.
The same model  was obtained later by Miles \&
 Salmon [5] through Hamilton's principle. 
Bazdenkov  {\it et al}. [6] derived the 2D GN system in the study of large-scale flows in planetary atmospheres and oceans.  Subsequently, Holm [7] demonstrated that it permits a Hamiltonian formulation by
introducing an appropriate noncanonical Lie-Poisson bracket. 
Recently, Lannes \& Bonneton [8] developed a systematic derivation of the various
asymptotic model equations including the 2D GN equations. They performed an asymptotic analysis of the Dirichlet-Neumann operator that originates from the formulation of
the water wave problem by Zakharov [9] and Craig \& Sulem [10]. See also Lannes's monograph (Lannes [11]) which details the derivation of various asymptotic model equations as well as their mathematical analysis.
In particular,  the well-posedness results of the GN models are presented in  Chapter 6 with some references on this topic. 
 \par
The purpose of the present paper is to generalize the 1D extended GN system mentioned above to the 2D system by making use of a novel asymptotic analysis, and show that it has the same 
Hamiltonian structure as that of the original 2D GN system.  
There are several new results in the current 2D extension. Among them, a highlight is an analysis of the linear dispersion relation for the extended GN equations
which makes it possible to explore the well-posedness of the linearized model equations. From the technical point of view, the 1D analysis can be applied to the 2D case as well. 
However, the computation involved becomes very complicated especially in the  vector formulation. \par
We consider the three-dimensional  irrotational flow of an incompressible and inviscid fluid of variable depth. 
 The effect of  surface tension is neglected since it has no appreciable influence on the current water wave phenomena. 
 It can be, however, incorporated in our formulation without difficulty.
The governing equation of the water wave problem is given in terms of  the dimensionless variables by
$$ \delta^2\nabla^2\phi+\phi_{zz}=0, \quad -1+\beta b<z<\epsilon\eta, \eqno(1.1)$$
$$\eta_t+\epsilon \nabla \phi{\Lcdot}\nabla\eta={1\over\delta^2} \phi_z, \quad z=\epsilon\eta,\eqno(1.2)$$
$$\phi_t+{\epsilon\over 2\delta^2}\left\{\delta^2(\nabla\phi)^2+\phi_z^2\right\}+\eta=0, \quad z=\epsilon\eta,\eqno(1.3)$$
$$\beta\delta^2\nabla b\Lcdot\nabla\phi=\phi_z, \quad z=-1+\beta b, \eqno(1.4)$$
subjected to the boundary conditions
$$\lim_{|{\bm x}|\rightarrow \infty}\nabla \phi({\bm x}, z, t)= {\bm 0},\quad \lim_{|{\bm x}|\rightarrow \infty}\phi_z({\bm x}, z, t)= 0, \quad
    -1+\beta b<z<\epsilon\eta, \quad \lim_{|{\bm x}|\rightarrow \infty}\eta({\bm x}, t)=0. \eqno(1.5)$$
Here, $\phi=\phi({\bm x}, z, t)$ is the velocity potential with ${\bm x}=(x, y)$ being a vector in the horizontal plane and $z$ the vertical coordinate pointing upwards, 
$\nabla =(\partial/\partial x, \partial/\partial y)$  is the 2D gradient operator, $\eta=\eta({\bm x},t)$ is the profile of the free surface, 
$b=b({\bm x})$ specifies the bottom topography,
and the subscripts $z$ and $t$ appended to $\phi$ and
 $\eta$ denote partial differentiations. 
 The notation $\nabla\phi\Lcdot\nabla\eta$ in (1.2) represents the scalar product between the 2D vectors $\nabla\phi$ and $\nabla\eta$.
 The equation $\eta=0$ represents the still water level.
 The boundary conditions (1.5) stem from the assumption that the fluid is at rest at infinity. We also assume that the total depth $h=h({\bm x}, t)$ of the fluid
 given by $h=1+\epsilon\eta-\beta b$ never vanishes. More precisely, $h$ is bounded from below by a positive constant. \par
The dimensional quantities, with tildes, are related to the corresponding dimensionless ones by the relations $\tilde {\bm x}=l{\bm x},
 \tilde z=h_0z, \tilde t=(l/c_0)t, \tilde\eta=a\eta$, $\tilde\phi=(gla/c_0)\phi$ and $\tilde b=b_0 b$, where $l$,  $h_0$, $a$, and $b_0$ 
 denote a characteristic wavelength, water depth,  wave amplitude and bottom profile, respectively. 
 $g$ is the acceleration due to the gravity, and $c_0=\sqrt{gh_0}$ is the long wave phase velocity.
 There arise the following three independent dimensionless parameters from the above scalings of the variables:
 $$\epsilon={a\over h_0},\quad \delta={h_0\over l},\quad \beta={b_0\over h_0}. \eqno(1.6)$$
 The nonlinearity parameter $\epsilon$  characterizes the magnitude of  nonlinearity
  whereas the dispersion parameter $\delta$  characterizes the dispersion or shallowness, and the parameter $\beta$ measures the variation of the bottom topography.  What is meant by "full nonlinearity" is that
  no restriction is imposed on the magnitude of $\epsilon$. Actually, $\epsilon$ is assumed to be of order 1 in our analysis.  On the other hand, we  impose the condition $\delta\ll 1$
  for the dispersion parameter which features the shallow water model equations.
  Note that the scaling $\epsilon/\delta^2=O(1)$ with
  $\epsilon, \delta \ll 1$ leads to the classical Boussinesq-type equations (Whitham [12]) whereas  the scaling $\epsilon/\delta=O(1)$ with
  $\epsilon, \delta \ll 1$ yields the Camassa-Holm and Degasperis-Procesi equations (Johnson [13], Constantin \& Lannes [14]). 
  A large number of works have been devoted to extend the Boussinesq-type equations by including both the higher-order dispersive and nonlinear effects.
  See the review papers written by  Kirby [15] and Madsen \& Sch\"affer [16, 17], for example. 
  \par
  The present paper is organized as follows. In \S 2, we reformulate the water wave problem posed by equations. (1.1)-(1.5) in terms of the total depth of fluid $h$ and the
  depth-averaged horizontal velocity $\bar {\bm u}$ which will be defined later. 
  It is noteworthy that the velocity at the free surface is determined completely  by the variables $h$ and $\bar {\bm u}$.
  The system of equations thus constructed consists of the  exact evolution equation for $h$ and an infinite-order Boussinesq-type equation for  $\bar {\bm u}$.  
  By truncating the latter equation at order $\delta^{2n}$, we obtain
  the extended GN equations which are accurate to $\delta^{2n}$, where $n$ is an arbitrary positive integer.  We call it the $\delta^{2n}$ model hereafter.
  The lowest-order approximation $n=1$ yields the  GN equations.
  We then derive the linear dispersion relation for the extended GN system, and investigate its characteristics in detail.
  We show that the Taylor series expansion of the linear dispersion relation for the $\delta^{2n}$ model coincides with that of the exact linearized water wave system up to order $\delta^{2n}$.
  The relevance of the models  in practical applications is also discussed while centering on the well-posedness of the linearized system of equations for the $\delta^{2n}$ model. 
In \S 3,  we derive, as illustrative examples, various approximate model equations which include the 2D $\delta^4$ model with a flat bottom topography
and the 2D $\delta^2$ model (or the GN model) with an uneven bottom topography. For the former model, the associated quantities such as
the linear dispersion relation, conservation laws and the pressure distribution at the fluid bottom are presented. Last, the 1D $\delta^6$ model
with a flat bottom topography is briefly described.  We emphasize that this model is able to  avoid the short wave instability which occurs in the $\delta^4$ model.
In \S4, we show that the extended GN equations can be formulated in a Hamiltonian form by introducing  an appropriate Lie-Poisson bracket as well as
the momentum density in place of $\bar {\bm u}$,
and they have the same Hamiltonian structure as that of the
 GN equations.
In \S 5, we demonstrate that the extended GN equations are equivalent to Zakharov's equations of motion for surface gravity waves.
This fact can be proved by rewriting the latter equations in terms of the momentum density instead of the velocity potential at the free surface.
Finally, \S 6 is devoted to conclusion where we summarize the main results and highlight some open problems.
In Appendix A, we prove the positivity of a polynomial associated with the linear dispersion relations for the $\delta^{2n}$ models  with odd $n$.
Appendix B provides a derivation of the pressure distribution at the fluid bottom while Appendix C gives a proof of the formula associated with the
variational derivative of the energy functional.
 \par
\bigskip
\noindent{\bf 2. Derivation of the extended Green-Naghdi equations}\par
\medskip
\noindent{(a) Extended GN system}\par
\medskip
\noindent The GN model is a system of equations for the  total depth of fluid $h$ and the
depth-averaged (or mean) horizontal velocity $\bar {\bm u}=(\bar u, \bar v)$. The latter variable is defined by
$$\bar {\bm u}={1\over h}\int^{\epsilon\eta}_{-1+\beta b}\nabla\phi({\bm x}, z, t){\mathrm d}z, \quad h=1+\epsilon\eta-\beta b, \eqno(2.1a)$$
with its components 
$$\bar u={1\over h}\int^{\epsilon\eta}_{-1+\beta b}\phi_x({\bm x}, z, t){\mathrm d}z, \quad \bar v={1\over h}\int^{\epsilon\eta}_{-1+\beta b}\phi_y({\bm x}, z, t){\mathrm d}z. \eqno(2.1b)$$
The horizontal component ${\bm u}=(u, v)$ and verical component $w$ of the surface velocity are given respectively by
$${\bm u}({\bm x}, t)=\nabla\phi({\bm x}, z, t)|_{z=\epsilon\eta}, \eqno(2.2a)$$
with its components
$$ u({\bm x}, t)=\phi_x({\bm x}, z, t)|_{z=\epsilon\eta}, \quad v({\bm x}, t)=\phi_y({\bm x}, z, t)|_{z=\epsilon\eta}, \eqno(2.2b)$$
and
$$w({\bm x}, t)=\phi_z({\bm x}, z, t)|_{z=\epsilon\eta}. \eqno(2.3)$$
\par
Let us now derive the  equations for $h$ and $\bm u$.  Since the procedure  of their derivation parallels  that of the 1D case (Matsuno [1]), we summarize the result. 
First, we multiply $(2.1a)$ by $h$ and then apply the divergence operator to the resultant expression.  This leads, after using (1.1) and (1.4), to 
$$w=\delta^2\{-\nabla\Lcdot(h\bar{\bm u})+\epsilon{\bm u}\Lcdot\nabla \eta\}.\eqno(2.4)$$
Insersion of $w$ from (2.4) into (1.2) now yields the evolution equation for $h=h({\bm x}, t)$:
$$h_t+\epsilon\nabla\Lcdot(h\bar{\bm u})=0. \eqno(2.5)$$
It is important that (2.5) is an {\it exact} equation without any approximation. \par
To obtain the equation of ${\bm u}$, we use the relation which follows from the definition of ${\bm u}$
$$\nabla\left(\phi_t|_{z=\epsilon\eta}\right)={\bm u}_t+\epsilon w_t\nabla \eta-\epsilon\eta_t\nabla w. \eqno(2.6)$$
Applying the gradient operator to (1.3) and using (2.6) as well as (2.4) and (2.5), we arrive at the  evolution equation for $\bm u$:
$${\bm u}_t+\epsilon w_t\nabla \eta+{\epsilon\over 2}\nabla {\bm u}^2+\epsilon^2({\bm u}\cdot\nabla \eta)\nabla w+\nabla\eta={\bf 0}. \eqno(2.7)$$
In the 1D case, equation (2.7) reduces to equation (2.8) of Matsuno [1]. \par
Now, we introduce the new quantity $\bm V$ by
$${\bm V}={\bm u}+\epsilon w\nabla \eta.\eqno(2.8)$$
We can rewrite equation (2.7) in terms of ${\bm V}$, giving
$${\bm V}_t+\epsilon\nabla\left({\bm u}\Lcdot{\bm V}-{1\over 2}{\bm u}^2-{1\over 2\delta^2}\,w^2+{\eta\over\epsilon}\right)={\bm 0}. \eqno(2.9)$$
\par
We  point out that a variant of equation (2.9) has been derived  by Witting [18]  to improve the 1D Boussinesq-type equations, and used by Dommermuth \& Yue [19] 
to develop an efficient numerical method for modelling gravity waves in the 2D setting.
The similar approaches have enabled us to improve the accuracy of
 the linear dispersion characteristics of the 2D Boussinesq-type equations (Kirby [15], Madsen \& Sch\"affer [16, 17]).
 Quite recently, equation (2.9) was generalized to the equation which is applicable to rotational flows (Castro \& Lannes [20]). 
 The latter equation was also derived in the framework of the Lagrangian formulation for the water wave problem (Gavrilyuk {\it et al}. [21]).
\par
Equation (2.9) represents an exact conservation law for the vector ${\bm V}$. To interpret the physical meaning of ${\bm V}$, we introduce the
velocity potential evaluated at the free surface
$$\psi({\bm x}, t)=\phi({\bm x}, \epsilon\eta,t). \eqno(2.10)$$
In view of the definition (2.2) and (2.3) of the surface velocity, the gradient of $\psi$ is found to be
$$\nabla\psi=(\nabla\phi+\epsilon\phi_z\nabla\eta)|_{z=\epsilon\eta}={\bm u}+\epsilon w\nabla \eta. \eqno(2.11)$$
It immediately follows from (2.8) and (2.11) that
$${\bm V}=\nabla\psi, \eqno(2.12)$$
implying that ${\bm V}$ is equal to the 2D gradient of the velocity potential evaluated at  the free surface, and it lies in the $(x, y)$ plane.
\par
We regard any 2D vector function ${\bm A}=(A_1, A_2)\in \mathbb{R}^2$ as  a function on $\mathbb{R}^3$ independent of the $z$ variable.
The rotation operator applied to the vector ${\bm A}$ is then defined by ${\rm rot}\,{\bm A}={\rm rot}\,({\bm A}, 0)^T=(0, 0, A_{2,x}-A_{1,y})^T$, showing that
it has only the $z$ component.  The similar definition will be used in \S 3 for the vector products like ${\bm A}\times{\bm B}$ and ${\bm A}\times({\bm B}\times{\bm C})$
with any vectors ${\bm A}, {\bm B}, {\bm C}\in \mathbb{R}^2$.
 For the potential flow under consideration, the rotation of ${\bm V}$ from (2.12) becomes zero identically, i.e.,
$${\rm rot}({\bm V})={\rm rot}(\nabla\psi)={\rm rot}\,(\psi_x, \psi_y, 0)^T=(0, 0, \psi_{yx}-\psi_{xy})^T\equiv {\bf 0}. \eqno(2.13)$$
\par
The system of equations (2.5) and (2.7) (or (2.9)) is  a consequence deduced from the basic Euler system (1.1)-(1.4). The extended GN equations
are obtained if one can express the variables ${\bm u}, w$ in equation (2.7) in terms of $h$ and $\bar{\bm u}$. As will be shown below, this
is always possible. Consequently, the evolution equation for $\bar{\bm u}$ can be recast in the form of an infinite-order Boussinesq-type equation
$${\bar{\bm u}}_t=\sum_{m=0}^\infty\delta^{2m}{\bm K}_m, \eqno(2.14)$$
where ${\bm K}_m \in \mathbb{R}^2$ are vector functions of $h$ and  $\nabla\Lcdot \bar{\bm u}, \nabla\Lcdot \bar{\bm u}_t$ as well as the spatial derivatives of these variables. 
  If one truncates the right-hand side of equation (2.14) at order $\delta^{2n}$, then equation (2.14)  yields the evolution equation for $\bar{\bm u}$
 which is accurate to $\delta^{2n}$ 
 $${\bar{\bm u}}_t=\sum_{m=0}^n\delta^{2m}{\bm K}_m.\eqno(2.15)$$
  The special case $n=1$ of equation (2.15) coupled with equation (2.5)  reduces to the original GN equations.
 In accordance with this fact, we call the system of equations (2.5) and (2.7) (or (2.9), (2.14)) with   $h$ and $\bar {\bm u}$ being the dependent variables  the extended GN system.
\par
\medskip
\noindent{(b) Expressions of the  velocities ${\bm u}, w$ and ${\bm V}$ in terms of $h$ and $\bar{\bm u}$}\par
\medskip
\noindent{(i) Flat bottom topography}\par
\noindent First, we solve the Laplace  equation (1.1) in the case of a flat bottom topography,  and express the surface velocity ${\bm u}, w$ and the  velocity ${\bm V}$ in terms of the variables $h$ and ${\bar{\bm u}}$.
This enables us to obtain a closed system of equations for the latter variables, i.e., the extended GN system.  Under the assumption $\delta^2\ll 1$ 
which is relevant to the shallow water models, the solution of equation (1.1) subjected to
the boundary condition (1.4) with $b=0$ can be written explicitly in the form of an infinite series
$$\phi({\bm x}, z, t)=\sum_{n=0}^\infty{(-1)^n\delta^{2n}\over (2n)!}\,(z+1)^{2n}\nabla^{2n}f, \eqno(2.16)$$
where $f=f({\bm x}, t)$ is the velocity potential at the fluid bottom.  See Whitam [12], Chapter 13 where  the similar formula is
presented for the 1D problem.
We substitute this expression into $(2.1a)$ and perform the integration with respect to $z$ to obtain
$$\bar{\bm u}=\nabla f+\sum_{n=1}^\infty{(-1)^n\delta^{2n}h^{2n}\over (2n+1)!}\,\nabla\nabla^{2n}f, \quad h=1+\epsilon\eta. \eqno(2.17)$$
 Using the formula $\nabla^2f=\nabla\Lcdot(\nabla f)$, we can rewrite (2.17) in an alternative form
$$\nabla f=\bar{\bm u}-\sum_{n=1}^\infty{(-1)^n\delta^{2n}h^{2n}\over (2n+1)!}\,\nabla\nabla^{2(n-1)}(\nabla\Lcdot \nabla f). \eqno(2.18)$$
\par
To derive the expansion of $\nabla f$ in terms of $h$ and $\bar{\bm u}$, we look for the solution in the form of an infinite series in $\delta^2$
$$\nabla f=\bar{\bm u}+\sum_{n=1}^\infty(-1)^n\delta^{2n}{\bm F}_n, \eqno(2.19)$$
where ${\bm F}_n\in \mathbb{R}^2$  are unknown vector functions to be determined below.
Substituting this expression into (2.18) and comparing the coefficients of $\delta^{2n}\ (n=1, 2, ...)$ on both sides, we obtain
$${\bm F}_1=-{h^2\over 6}\,\nabla(\nabla\Lcdot\bar{\bm u}), \eqno(2.20a)$$
$${\bm F}_n=-{h^{2n}\over (2n+1)!}\,\nabla\nabla^{2(n-1)}(\nabla\Lcdot\bar{\bm u})
-\sum_{r=1}^{n-1}{h^{2r}\over (2r+1)!}\,\nabla\nabla^{2(r-1)}(\nabla\Lcdot{\bm F}_{n-r}),\quad (n\geqslant 2). \eqno(2.20b)$$
The recursion relation $(2.20b)$ for ${\bm F}_n$ can be solved successively with the initial condition $(2.20a)$, the first two of which read
$${\bm F}_2=-{h^4\over 120}\nabla\nabla^{2}(\nabla\Lcdot\bar{\bm u})+{h^2\over 36}\nabla\nabla\Lcdot\{h^2\nabla(\nabla\Lcdot\bar{\bm u})\}, \eqno(2.21a)$$
$${\bm F}_3=-{h^6\over 5040}\nabla\nabla^{4}(\nabla\Lcdot\bar{\bm u})-{h^2\over 6}\nabla(\nabla\Lcdot{\bm F}_2)-{h^4\over 120}\nabla\nabla^2(\nabla\Lcdot{\bm F}_1). \eqno(2.21b)$$
\par
The series expansions of  ${\bm u}, w$ and  ${\bm V}$  can be derived simply by substituting (2.19) with ${\bm F}_n$ from (2.20) and (2.21)
into (2.2), (2.3) and (2.12), respectively. We write them up to order $\delta^4$ for later use:
$${\bm u}=\nabla f+\sum_{n=1}^\infty{(-1)^n\delta^{2n}h^{2n}\over (2n)!}\,\nabla\nabla^{2(n-1)}(\nabla\Lcdot \nabla f)$$
$$=\bar{\bm u}-{\delta^2\over 3}\,h^2\nabla(\nabla\Lcdot\bar{\bm u})+\delta^4\left[-{1\over 18}h^2\nabla\nabla\Lcdot\{h^2\nabla(\nabla\Lcdot\bar{\bm u})\}+{1\over 30}h^4\nabla\nabla^2(\nabla\Lcdot\bar{\bm u})\right]
+O(\delta^6), \eqno(2.22)$$
$$w=\sum_{n=1}^\infty{(-1)^n\delta^{2n}h^{2n-1}\over (2n-1)!}\,\nabla^{2n}f
=-\delta^2h\nabla\Lcdot\bar{\bm u}-{\delta^4\over 3}\,h^2\nabla h\Lcdot \nabla(\nabla\Lcdot\bar{\bm u})+O(\delta^6), \eqno(2.23)$$
$${\bm V}=\nabla f+\sum_{n=1}^\infty{(-1)^n\delta^{2n}\over (2n)!}\nabla(h^{2n}\nabla^{2(n-1)}\nabla^2f)$$
$$=\bar{\bm u}-{\delta^2\over 3h}\,\nabla(h^3\nabla\Lcdot\bar{\bm u})-{\delta^4\over 45h}\,\nabla[\nabla\Lcdot\{h^5\nabla(\nabla\Lcdot\bar{\bm u})\}]+O(\delta^6). \eqno(2.24)$$
Substituting (2.24) into (2.13), we can express ${\rm rot}(\bar{\bm u})$ as a function of $h$ and $\nabla\Lcdot\bar{\bm u}$, the
explict form of which will be presented in \S 3 in deriving the evolution equation for $\bar{\bm u}$. \par
\medskip
\noindent{(ii)  Uneven bottom topography}\par
\noindent The effect of an uneven bottom topography on the propagation characteristics of water waves is prominent in the coastal zone.
Here, we provide the formulas of ${\bm u}, w$ and ${\bm V}$ in terms of $h, \bar{\bm u}$ and $b$.  In this case, the solution of the Laplace equation (1.1) subjected to the
boundary condition (1.5) can be written in the form
$$\phi({\bm x}, z, t)=\sum_{n=0}^\infty(z+1-\beta b)^n\phi_n({\bm x}, t), \eqno(2.25)$$
where  the orders of  unknown functions $\phi_n$ are to be determined. See, Yoon \& Liu [22] for example, where a similar manipulation has been performed.
Substituting (2.25) into equation (1.1), we obtain the recursion relation among $\phi_n$,  $\phi_{n+1}$ and $\phi_{n+2}$
$$\phi_{n+2}=-{\delta^2[\nabla^2\phi_n-2\beta(n+1)\nabla b\Lcdot \nabla \phi_{n+1}-\beta (n+1)(\nabla^2b)\phi_{n+1}\over
(n+1)(n+2)\{1+\beta^2\delta^2(\nabla b)^2\}}, \quad n \geqslant 0. \eqno(2.26)$$
On the other hand, it follows from (1.4) and (2.25) that
$$\phi_1={\beta\delta^2\nabla b\Lcdot\nabla \phi_0\over 1+\beta^2\delta^2(\nabla b)^2}. \eqno(2.27)$$
Thus, the velocity potential is expressed in terms of a single unknown variable $\phi_0$, i.e., the velocity potential at the fluid bottom which corresponds to $f$ in (2.16).
For small $\delta^2$, the first three of $\phi_n$ are found to be
$$\phi_1=\beta\delta^2(\nabla b\Lcdot\nabla \phi_0)\{1-\beta^2\delta^2(\nabla b)^2\}+O(\delta^6), \eqno(2.28a)$$
$$\phi_2=-{\delta^2\over 2}\nabla^2\phi_0+\beta^2\delta^4\left\{{1\over 2}(\nabla b)^2\nabla^2\phi_0+\nabla b\Lcdot \nabla(\nabla b\Lcdot \nabla\phi_0)
+{1\over 2}\nabla^2b(\nabla b\Lcdot \nabla\phi_0)\right\}+O(\delta^6), \eqno(2.28b)$$
$$\phi_3=-{\beta\delta^4\over 6}\{\nabla^2(\nabla b\Lcdot \nabla\phi_0)+2\nabla b\Lcdot\nabla(\nabla^2\phi_0)+\nabla^2b\nabla^2\phi_0\}+O(\delta^6). \eqno(2.28c)$$
\par
The depth-averaged horizontal velocity $\bar{\bm u}$ and the surface velocity $({\bm u}, w)$ are expressed in terms of $\phi_n$ by introducing (2.25) into (2.1),  (2.2) and (2.3). 
 Explicitly, they read
$$\bar {\bm u}=\sum_{n=0}^\infty{h^n\over n+1}\,\nabla\phi_n-\beta\nabla b\sum_{n=1}^\infty h^{n-1}\phi_n, \quad h=1+\epsilon\eta-\beta b, \eqno(2.29)$$
$${\bm u}=\sum_{n=0}^\infty h^n\,\nabla\phi_n-\beta\nabla b\sum_{n=1}^\infty n h^{n-1}\phi_n, \eqno(2.30)$$
$$w=\sum_{n=1}^\infty nh^{n-1}\phi_n. \eqno(2.31)$$
Inverting (2.29) and using (2.28), we can express $\nabla\phi_0$ in terms of $\bar{\bm u}, h$ and $b$. The approximate expression 
which retains the terms of order $\delta^2$ is given by 
$$\nabla\phi_0=\bar{\bm u}+\delta^2\left[{h^2\over 6}\nabla(\nabla\Lcdot\bar{\bm u})-{\beta\over 2}\big\{h\nabla(\nabla b\Lcdot\bar{\bm u})+(h\nabla\Lcdot\bar{\bm u})\nabla b\}
+\beta^2(\nabla b\Lcdot\bar{\bm u})\nabla b\right]+O(\delta^4). \eqno(2.32)$$
Substitution of (2.28) with (2.32) into (2.30) and (2.31) yields the approximate expressions of ${\bm u}$ and $w$ 
$${\bm u}=\bar{\bm u}+\delta^2\left[-{h^2\over 3}\nabla(\nabla\Lcdot\bar{\bm u})+{\beta\over 2}\big\{h\nabla(\nabla b\Lcdot\bar{\bm u})+(h\nabla\Lcdot\bar{\bm u})\nabla b\big\}\right]+O(\delta^4), \eqno(2.33)$$
$$w=\delta^2(-h\nabla\Lcdot\bar{\bm u}+\beta\nabla b\Lcdot\bar{\bm u})+O(\delta^4). \eqno(2.34)$$
 Last, by making use of (2.33) and (2.34),  ${\bm V}$ from (2.8) is shown to have an approximate expression
$${\bm V}=\bar{\bm u}+{\delta^2\over h}\left[-{1\over 3}\nabla(h^3\nabla \Lcdot\bar{\bm u})+{\beta\over 2}\{\nabla(h^2\nabla b\Lcdot\bar{\bm u})-h^2\nabla b(\nabla\Lcdot\bar{\bm u})\}
+\beta^2h\nabla b(\nabla b\Lcdot\bar{\bm u})\right]+O(\delta^4). \eqno(2.35)$$
\par
\medskip
\noindent{(c) Linear dispersion relation for the extended GN system} \par
\noindent Here, we show that the exact linear dispersion relation for the current water wave problem can be derived  from the extended GN system, and discuss its structure. 
We consider the flat bottom case for simplicity.  
Linearization of  equations (2.5) and (2.7) about the 
uniform state $h=1$ and $\bar{\bm u}={\bf 0}$ yields the system of linear equations  for $\eta$ and $\bar{\bm u}$
$$\eta_t+\nabla\Lcdot \bar{\bm u}=0, \quad {\bm u}_t+\nabla\eta={\bm 0}. \eqno(2.36)$$
We  eliminate the variable $\eta$ from the system of equations (2.36)  and obtain the
linear wave equation for $\bar{\bm u}$
$${\bm u}_{tt}-\nabla(\nabla\Lcdot \bar{\bm u})={\bf 0}. \eqno(2.37)$$
Recall that the variable ${\bm u}$ is a linear function of $\bar{\bm u}$ and its spatial derivatives.  
It follows from  (2.22)  with $h=1$  that
$${\bm u}=\nabla f+\sum_{n=1}^\infty{(-1)^n\delta^{2n}\over (2n)!}\,\nabla\nabla^{2(n-1)}(\nabla\Lcdot \nabla f), \eqno(2.38)$$
where $\nabla f$ is given by (2.19). As inspected from (2.20) and (2.21) with $h=1$, we can put $F_n$ in the form
$$ {\bm F}_n=\alpha_n \nabla\nabla^{2(n-1)}(\nabla\Lcdot\bar{\bm u}), \quad n\geqslant 1, \eqno(2.39)$$
where $\alpha_n$ are unknown constants. Substituting (2.39) into $(2.20b)$ with $h=1$, we obtain  the recursion relation for $\alpha_n$
$$\alpha_1=-{1\over 6}, \quad\alpha_n=-{1\over (2n+1)!}-\sum_{r=1}^{n-1}{\alpha_{n-r}\over (2r+1)!},\quad n\geqslant 2. \eqno(2.40)$$
The expression of  ${\bm u}$ in terms of $\bar{\bm u}$ follows by
inserting (2.19) with (2.39) into (2.38), giving
$${\bm u}=\bar{\bm u}+\sum_{n=1}^\infty(-1)^n\delta^{2n}\left\{{1\over (2n)!}+\sum_{r=0}^{n-1}{\alpha_{n-r}\over (2r)!}\right\}\nabla\nabla^{2(n-1)}(\nabla\Lcdot\bar{\bm u}). \eqno(2.41)$$
\par
In order to examine the linear dispersion characteristics of equation (2.37) with ${\bm u}$ from (2.41), we assume the solution of the form
$\bar{\bm u}=\bar{\bm u}_0\,{\rm e}^{{\rm i}({\bm k}\cdot{\bm x}-\omega t)}$
 where $\bar{\bm u}_0$ is a 2D constant vector,  ${\bm k}$ is the 2D wavenumber vector and $\omega$ is the angular frequency.
To proceed, we first take the divergence of equation (2.37) and then replace ${\bm u}$ by (2.41).  Last, we 
substitute $\bar{\bm u}$ given above into the resultant equation for $\bar{\bm u}$
and find that the linear dispersion relation takes the form
$$\omega^2={k^2\over D(k\delta)}, \quad (k=|{\bm k}|), \eqno(2.42a)$$
where 
$$D(k\delta)=1+\sum_{n=1}^\infty(k\delta)^{2n}\left\{{1\over (2n)!}+\sum_{r=0}^{n-1}{\alpha_{n-r} \over (2r)!}\right\}. \eqno(2.42b)$$
Now, in view of the recursion relation (2.40), we deduce
$$\sum_{n=1}^\infty(k\delta)^{2n}\alpha_n=-{1\over k\delta}(\sinh\,k\delta-k\delta)\left\{1+\sum_{n=1}^\infty(k\delta)^{2n}\alpha_n\right\}, \eqno(2.43)$$
from which we obtain
$$\sum_{n=1}^\infty(k\delta)^{2n}\alpha_n=-{\sinh\,k\delta-k\delta\over \sinh\,k\delta}. \eqno(2.44)$$
The similar calculation yields, after using (2.44),  the relation
$$\sum_{n=1}^\infty(k\delta)^{2n}\sum_{r=0}^{n-1}{\alpha_{n-r} \over (2r)!}=\sum_{r=0}^\infty{(k\delta)^{2r}\over (2r)!}\sum_{n=1}^\infty(k\delta)^{2n}\alpha_n
=-(\sinh\,k\delta-k\delta)\coth\,k\delta. \eqno(2.45)$$
It follows from $(2.42b)$ and (2.45) that
$$D(k\delta)=k\delta\,\coth\,k\delta,\eqno(2.46)$$
which, substituted into $(2.42a)$, leads to the linear dispersion relation for the extended GN system
$$\omega^2={k\over\delta}\tanh\,k\delta. \eqno(2.47)$$   
The above expression coincides perfectly with that derived from the linearized system of equations for the current water wave problem (1.1)-(1.5).
\par
The $\delta^{2n}$ GN model incorporates the dispersive terms of order $\delta^{2n}$. Referring to equations (2.5) and (2.15), 
one can write it in the form 
$$h_t+\epsilon\nabla\Lcdot(h\bar{\bm u})=0, \quad {\bar{\bm u}}_t=\sum_{m=0}^n\delta^{2m}{\bm K}_m. \eqno(2.48)$$
The linear dispersion relation of the above system is given by (2.42) with $D$ truncated at order $(k\delta)^{2n}$.
 To detail the dispersion characteristics of this model, we
introduce the function $D_{2n}(\kappa)$ by
$$D_{2n}(\kappa)=1+\sum_{r=1}^n{(-1)^{r-1}2^{2r}\over (2r)!}B_r\kappa^{2r}, \eqno(2.49a)$$
where $B_r$ are Bernoulli's numbers defined by
$$B_r={2(2r)!\over(2\pi)^{2r}}\sum_{j=1}^\infty{1\over j^{2r}}, \quad r\geqslant 1,\eqno(2.49b)$$
the first four of which are given by $B_1=1/6, B_2=1/30, B_3=1/42, B_4=1/30$. 
Explicitly, the first three of $D_{2n}$ read
$$D_2=1+{\kappa^2\over3}, \quad D_4=1+{\kappa^2\over3}-{\kappa^4\over 45}, \quad D_6=1+{\kappa^2\over3}-{\kappa^4\over 45}+{2\kappa^6\over 945}. \eqno(2.50)$$
\par
In view of the fact that 
 $D_{2n}$ is a polynomial of order $2n$ in $\kappa$ and coincides with the Taylor series expansion of the
function $\kappa\coth\,\kappa$ truncated at the same order (see Abramowitz \&  Stegun [23], Chapters 4 and 23),
the linear dispersion relation for the $\delta^{2n}$  model (2.48) is represented by
$$\omega^2={k^2\over D_{2n}(k\delta)}. \eqno(2.51)$$
\par
Using the inequality for the Bernoulli numbers,  
we can show that $D_{2n}$ with odd $n$ are  positive for all $k\delta$.  More precisely, they have a lower bound  1. As a result,  an estimate $\omega/k\leqslant 1$  for $k\delta\geqslant 0$ follows.
  We provide a proof of this statement in Appendix A. Consequently, 
the $\delta^{2n}$ models with odd $n$ have a nice property as long as the linear dispersion characteristic is concerned.
Actually, they have smooth dispersion relations without any singularities,
 and possess an important feature that the exact linear dispersion relation has, i.e., $\omega/k=\sqrt{\tanh\,k\delta/k\delta}\leqslant 1$ for $k\delta\geqslant 0$.
On the other hand, $D_{2n}$ models with even $n$ exhibit single positive zero. 
 For example, the positive zeros of $D_4$, $D_8$ and $D_{10}$ are found to be
$4.19, 3.63$ and $3.33$, respectively.  An asymptotic analysis shows that the zero of $D_{2n}$ with even $n$ approaches a constant value $\pi$ as $n$ tends to infinity.
These results imply that $\omega$ from (2.51) has a singularity and becomes pure imaginary for values of $k\delta$ exceeding the zero.
It turns out that the $\delta^{2n}$ models with  even $n$ exhibit an unphysical dispersion characteristic which
leads to the ill-posedness result for the linearized systems of equations, and 
 may cause  instabilities  in short wave solutions in practical numerical computations.
In accordance with these observations, the $\delta^{2n}$ models with odd $n$ 
may be more tractable as the practical model equations than the  $\delta^{2n}$ models with  even $n$.
Although this important issue should be investigated further from both theoretical and numerical points of view, we leave it for a future work. 
\par
\bigskip
\noindent{\bf 3. Approximate model equations}\par
\medskip
\noindent The method developed in \S 2 enables us to extend systematically the GN equations  to include the arbitrary order of dispersion.
In this section, we derive, as an illustrative example, an extended GN model ($\delta^4$ model) which is accurate to order $\delta^4$ in the case of a flat bottom topography.
 Subsequently, we obtain the linear dispersion relation and the conservation laws for the model. We also evaluate the pressure 
distribution at the fluid bottom. \par
In the case of an uneven bottom topography, the derivation of the $\delta^4$ model  can be  performed without difficulty. However, since the expression of the $\delta^4$ model is
too complicated to write down, we address only the  $\delta^2$ model or GN  model, and confirm that it reproduces the corresponding model equation derived by different methods.
\par
\medskip
\noindent {(a) The $\delta^4$ model}\par
\medskip
\noindent{(i)  Derivation of the $\delta^4$ model with a flat bottom topography}\par
\noindent For the purpose of deriving the $\delta^4$ model with  a flat bottom topography,
we only need  the evolution equation for $\bar{\bm u}$ since the equation for $h$  is already at hand, as indicated by  equation (2.5). 
The procedure for obtaining the equation for $\bar{\bm u}$ can be performed straightforwardly and hence we outline the derivation. \par
We start from equation (2.9) for ${\bm V}$.  The second term on the left-hand side of  equation (2.9) is expressed in terms of $h$ and $\bar{\bm u}$ by
substituting (2.22)-(2.24) into the corresponding terms.  Retaining terms up to order $\delta^4$, we obtain
$${\bm u}\Lcdot{\bm V}-{1\over 2}{\bm u}^2-{1\over 2\delta^2}\,w^2={1\over 2}\,\bar{\bm u}^2
-\delta^2\left\{{1\over 3h}\,\bar{\bm u}\Lcdot\nabla(h^3\nabla\cdot\bar{\bm u})+{1\over 2}(h\nabla\Lcdot\bar{\bm u})^2\right\}$$
$$+\delta^4\left[{h^4\over 18}\{\nabla(\nabla\Lcdot\bar{\bm u})\}^2-{1\over 45h}\bar{\bm u}\Lcdot\nabla\Bigl\{\nabla \Lcdot(h^5\nabla(\nabla\Lcdot\bar{\bm u}))\Bigr\}\right]+O(\delta^6). \eqno(3.1)$$
In the process of deriving the evolution equation for $\bar{\bm u}$,  one needs to evaluate the term $\nabla(\bar{\bm u}^2)$.  By virtue of the
well-known vectorial identity, it can be rewritten as
${1\over 2}\nabla{\bar{\bm u}}^2=(\bar{\bm u}\Lcdot\nabla){\bar{\bm u}}+\bar{\bm u}\times{\rm rot}(\bar{\bm u})$, where
${\rm rot}(\bar{\bm u})$ is computed by substituting ${\bm V}$ from (2.24) into the relation (2.13) to give
$${\rm rot}(\bar{\bm u})={\delta^2\over 3}\,{\rm rot}\left[{1\over h}\nabla(h^3\nabla\Lcdot\bar{\bm u})\right]
+{\delta^4\over 45}\,{\rm rot}\left[{1\over h}\nabla\Bigl\{\nabla\Lcdot(h^5\nabla(\nabla\Lcdot\bar{\bm u}))\Bigr\}\right]+O(\delta^6). \eqno(3.2)$$
\par
 Now, we introduce (2.24) and (3.1) into equation (2.9) and use the expression (3.2). 
The time differentiation $h_t$ which stems from ${\bm V}_t$ is replaced by $-\epsilon\nabla\cdot(h\bar{\bm u})$ in accordance with equation (2.5).
Rearranging terms, we finally arrive at the evolution equation for $\bar{\bm u}$:
$${\bar{\bm u}}_t+\epsilon(\bar{\bm u}\Lcdot\nabla){\bar{\bm u}}+\nabla\eta=\delta^2R_1+\delta^4R_2+O(\delta^6), \eqno(3.3a)$$
with
$$R_1={1\over 3h}\nabla\Bigl[h^3\{\nabla\Lcdot{\bar{\bm u}}_t+\epsilon(\bar{\bm u}\Lcdot\nabla)(\nabla\Lcdot\bar{\bm u})-\epsilon(\nabla\Lcdot\bar{\bm u})^2\}\Bigr], \eqno(3.3b)$$
$$R_2={1\over 45h}\nabla\Bigl[\nabla\Lcdot\bigl\{h^5\nabla(\nabla\Lcdot{\bar{\bm u}}_t)+\epsilon h^5(\nabla^2(\nabla\Lcdot\bar{\bm u}))\bar{\bm u}-5\epsilon h^5(\nabla\Lcdot\bar{\bm u})\nabla(\nabla\Lcdot\bar{\bm u})
+\epsilon\nabla h^5\times(\bar{\bm u}\times\nabla(\nabla\Lcdot\bar{\bm u}))\bigr\}$$
$$-2\epsilon h^5\{\nabla(\nabla\Lcdot\bar{\bm u})\}^2\Bigr]
-{\epsilon\over 45h}\Bigl[\nabla\Lcdot\{h^5\nabla(\nabla\Lcdot\bar{\bm u})\}\nabla(\nabla\Lcdot\bar{\bm u})+{h^5\over2}\nabla\{\nabla(\nabla\Lcdot\bar{\bm u})\}^2\Bigr]. \eqno(3.3c)$$
Note that the term represented by the vector triple product in $R_2$ is a vector in the $(x, y)$ plane.   
It is important that the second term of $R_2$ multiplied by $h$  can be recast in a conservation form. Namely,
$$\nabla\Lcdot\{h^5\nabla(\nabla\Lcdot\bar{\bm u})\}\nabla(\nabla\Lcdot\bar{\bm u})+{h^5\over2}\nabla\{\nabla(\nabla\Lcdot\bar{\bm u})\}^2
=\sum_{j=1}^2{\partial\over\partial x_j}\left[h^5{\partial (\nabla\Lcdot\bar{\bm u})\over\partial x_j}\,\nabla(\nabla\Lcdot\bar{\bm u})\right], \eqno(3.4)$$
where we have put $x_1=x$ and $x_2=y$. 
\par
Various reductions are possible for the $\delta^4$ model.
Indeed, if we neglect the $\delta^4$ terms in equation (3.3), then it reduces to the 2D GN system when coupled with equation (2.5)
$$h_t+\epsilon\nabla\Lcdot(h\bar{\bm u})=0, \eqno(3.5a)$$
$${\bar{\bm u}}_t+\epsilon(\bar{\bm u}\Lcdot\nabla){\bar{\bm u}}+\nabla\eta
={\delta^2\over 3h}\nabla\left[h^3\{\nabla\Lcdot{\bar{\bm u}}_t+\epsilon(\bar{\bm u}\Lcdot\nabla)(\nabla\Lcdot\bar{\bm u})-\epsilon(\nabla\Lcdot\bar{\bm u})^2\}\right], \eqno(3.5b)$$
whereas the  $\delta^4$ model reduces to the classical 2D Boussinesq system
$$h_t+\epsilon\nabla\Lcdot(h\bar{\bm u})=0, \eqno(3.6a)$$
$${\bar{\bm u}}_t+\epsilon(\bar{\bm u}\Lcdot\nabla){\bar{\bm u}}+\nabla\eta={\delta^2\over 3}\nabla(\nabla\Lcdot\bar{\bm u}_t), \eqno(3.6b)$$
after neglecting the $\epsilon\delta^2$ and higher-order terms.
On the other hand, the 1D forms of equations (2.5) and (3.3) become
$$h_t+\epsilon(h\bar u)_x=0, \eqno(3.7a)$$
$$\bar u_t+\epsilon\bar u\bar u_x+\eta_x={\delta^2\over 3h}\left\{h^3(\bar u_{xt}+\epsilon\bar u\bar u_{xx}-\epsilon\bar u_x^2)\right\}_x$$
$$+{\delta^4\over 45h}\left[\left\{h^5(\bar u_{xxt}+\epsilon\bar u\bar u_{xxx}-5\epsilon \bar u_x\bar u_{xx})\right\}_x-3\epsilon h^5\bar u_{xx}^2\right]_x+O(\delta^6), \eqno(3.7b)$$
which are in agreement with equations (2.5) and (2.21) of Matsuno [1], respectively. \par
\bigskip
\noindent{(ii)  Linear dispersion relation}\par
\noindent The system of equations (2.5) and (3.3) linearized about the uniform state $h=1$ and $\bar{\bm u}={\bf 0}$
is simply expressed  as
$$\eta_t+\nabla\Lcdot\bar{\bm u}={\bm 0}, \quad {\bar{\bm u}}_t+\nabla\eta={\delta^2\over 3}\nabla(\nabla\Lcdot{\bar{\bm u}}_t)+{\delta^4\over 45}\nabla\nabla^2(\nabla\Lcdot{\bar{\bm u}}_t). \eqno(3.8)$$
The linear dispersion relation for the system (3.8)  is given by
 $$\omega^2={k^2\over 1+{1\over 3}(k\delta)^2-{1\over 45}(k\delta)^4}, \quad k=|{\bm k}|. \eqno(3.9)$$
 The property of (3.9) has been discussed in the 1D case. See Matsuno [1].
  Note that $\omega$ from (3.9) exhibits a singularity at $k\delta\simeq 4.19$.
As already discussed in \S 2(c), this shortcoming can be overcome if one employs
 the $\delta^6$ model, for example.  See \S 3(c).\par
 \medskip
\noindent{(iii) Conservation laws}\par
\noindent The $\delta^4$ model derived here exhibits the following four conservation laws:
$$M=\int_{\mathbb{R}^2}(h-1){\mathrm d}{\bm x}, \eqno(3.10)$$
$${\bm P}=\int_{\mathbb{R}^2}h{\bar{\bm u}}\,{\mathrm d}{\bm x}, \eqno(3.11)$$
$$H={\epsilon^2\over 2}\int_{\mathbb{R}^2}\left[h{\bar{\bm u}}^2+{\delta^2\over 3}h^3(\nabla\Lcdot\bar{\bm u})^2
-{\delta^4\over 45}h^5\{\nabla(\nabla\Lcdot\bar{\bm u})\}^2+{1\over\epsilon^2}(h-1)^2\right]{\mathrm d}{\bm x}, \eqno(3.12)$$
$${\bm L}=\epsilon\int_{\mathbb{R}^2}\left[\bar{\bm u}-{\delta^2\over 3h}\nabla(h^3\nabla\Lcdot\bar{\bm u})
-{\delta^4\over 45 h}\nabla\left[\nabla\Lcdot\{h^5\nabla(\nabla\Lcdot\bar{\bm u}\}\right]\right]{\mathrm d}{\bm x}, \eqno(3.13)$$
where we have used the notation $\int_{\mathbb{R}^2}F({\bm x}, t){\mathrm d}{\bm x}=\int^{\infty}_{-\infty}\int^{\infty}_{-\infty}F({\bm x}, t){\mathrm d}x{\mathrm d}y$ for any function $F$ decreasing rapidly at infinity.  The factors $\epsilon^2$ 
and $\epsilon$ attached in front of the integral sign in $H$ and ${\bm L}$, respectively are only for convenience. 
The quantities $M, {\bm P}$ and $H$ represent the conservation of the mass, momentum and total energy, respectively, which can be confirmed directly by using equations (2.5) and (3.3).
The fourth conservation law ${\bm L}$ follows from (2.9) and (2.24). The geometrical interpretation of ${\bm L}$ has been discussed in detail in the 1D case. See {\bf Remark 6} of Matsuno [1].
This quantity will appear in \S 4 in developing the Hamiltonian formulation of the 2D extended GN system. \par
\medskip
\noindent{(iv) Pressure distribution at the fluid bottom}\par
\noindent The pressure distribution at the flat bottom in the context of nonlinear shallow-water waves has attracted considerable attention. See Constantin  {\it et al}. [24] and Deconinck  {\it et al}. [25].
Recently,  the calculation of the bottom pressure due to the passage of a large amplitude solitary wave was carried out in the framework
of the 1D GN model, and the results were compared with those obtained by the linear theory (Touble \& Pelinovsky [26], Pelinovsky  {\it et al}. [27]). 
Here, we present  an explicit formula for the bottom pressure $P_b$ in terms of $h$ and $\bar{\bm u}$.  The derivation of the formula is given in Appendix B. \par
Now, the bottom pressure  can be expressed in the form
$$P_b=h+\delta^2P_{b1}+\delta^4P_{b2}+O(\delta^6), \eqno(3.14a)$$
with
$$P_{b1}=-{\epsilon h^2\over 2}\Big[\nabla\Lcdot{\bar{\bm u}}_t+\epsilon \bar{\bm u}\Lcdot\nabla(\nabla\Lcdot\bar{\bm u})-\epsilon(\nabla\Lcdot\bar{\bm u})^2\Big],\eqno(3.14b)$$
$$P_{b2}=-{\epsilon\over 24}\nabla\Lcdot\Big[h^4\{\nabla({\bar{\bm u}}_t)+\epsilon\nabla^2(\nabla\Lcdot\bar{\bm u})\bar{\bm u}-5\epsilon(\nabla\Lcdot\bar{\bm u})\nabla(\nabla\Lcdot\bar{\bm u})\}$$
$$+4\epsilon h^3\{(\nabla h\Lcdot \bar{\bm u})\nabla(\nabla\Lcdot\bar{\bm u})-(\nabla(\nabla\Lcdot\bar{\bm u})\Lcdot\nabla h)\bar{\bm u}\}\Big]. \eqno(3.14c)$$
\par
Various simplifications are possible for the above formula. Among them, we consider the two special cases. The first example is the bottom pressure caused by the passage of a travelling wave of the
form $h=h(\sigma), \bar{\bm u}=\bar{\bm u}(\sigma)$ with $\sigma={\bm k}\Lcdot {\bm x}-\omega t$  being the phase variable.  Substituting these forms into equation (2.5) and integrating it 
with respect to $\sigma$ under the
boundary conditions $h\rightarrow 1, \bar{\bm u}\rightarrow {\bm 0}$ as $|\sigma|\rightarrow \infty$, we obtain
$${\bm k}\Lcdot{\bar{\bm u}}={\omega\over\epsilon h}\left(1-{1\over h}\right). \eqno(3.15)$$
This expression enables us to represent $P_b$ from (3.14) in terms of the total fluid depth $h$.
 Actually,  replacing the time derivative $\partial/\partial t$ and the gradient operator $\nabla $ by $-\omega\,d/d\sigma$ and ${\bm k}\,d/d\sigma$, respectively and  then  using (3.15),  we find a compact expression for $P_b$
$$P_b=h+{\omega^2\delta^2\over 2}\left({h^\prime\over h}\right)^\prime+{\omega^2k^2\delta^4\over 24}\left[{1\over h}\left(h^2h^{\prime\prime\prime}-hh^\prime h^{\prime\prime}-4{h^\prime}^3\right)\right]^\prime, \eqno(3.16)$$
where the prime refers to differentiation with respect to $\sigma$.  \par
The second example is the 1D reduction of (3.14).  The corresponding formula can be written simply in the form
$$P_b=h-{\epsilon\delta^2\over 2}h^2(\bar u_{xt}+\epsilon \bar u\bar u_{xx}-\epsilon{\bar u_x}^2)
-{\epsilon\delta^4\over 24}\Big[h^4(\bar u_{xxt}+\epsilon \bar u\bar u_{xxx}-5\epsilon \bar u_x\bar u_{xx})\Big]_x. \eqno(3.17)$$
Within the framework of the GN model, the last term on the right-hand side of (3.17) is absent. It was used recently to evaluate the bottom pressure caused
by the passage of a soliton, and the resultant expression was compared with the formula derived from the linear theory (Pelinovsky {\it  et al}. [27]). \par
\medskip
\noindent{(b) The  GN model with an uneven bottom topography} \par
\medskip
\noindent In accordance with the method developed in \S 2,  let us derive the  GN model which takes into account  an uneven bottom topography.
Since its derivation  is almost parallel to that of the flat bottom case, we describe only the outline. \par
The expression corresponding to (3.1) follows by using (2.33)-(2.35), giving
$${\bm u}\Lcdot{\bm V}-{1\over 2}{\bm u}^2-{1\over 2\delta^2}\,w^2={1\over 2}\,\bar{\bm u}^2
+\delta^2\bigg[-{1\over 3h}\,\bar{\bm u}\Lcdot\nabla(h^3\nabla\Lcdot\bar{\bm u})-{1\over 2}(h\nabla\Lcdot\bar{\bm u})^2$$
$$+\beta\left\{{1\over 2}h\nabla\Lcdot((\nabla b\Lcdot\bar{\bm u})\bar{\bm u})+(\bar{\bm u}\Lcdot\nabla h)\nabla b\Lcdot\bar{\bm u}\right\}
+{1\over 2}\beta^2(\nabla b\Lcdot\bar{\bm u})^2\bigg]+O(\delta^4). \eqno(3.18)$$
\par
Next, introducing (2.35) into the relation ${\rm rot}({\bm V})={\bf 0}$ from (2.13), we obtain
$${\rm rot}(\bar{\bm u})=\delta^2\bigg[-{h\over 3}\nabla h\times\nabla(\nabla\Lcdot\bar{\bm u})+{\beta\over 2}\{\nabla h\times\nabla(\nabla b\Lcdot\bar{\bm u})
+\nabla(h\nabla\Lcdot\bar{\bm u})\times\nabla b\}$$
$$-\beta^2\nabla(\nabla b\Lcdot \bar{\bm u})\times\nabla b\bigg]+O(\delta^4). \eqno(3.19)$$
\par
Last, we substitute (2.35) and (3.18) into equation (2.9) and use the formula (3.19) whereby we replace $h_t$ by $-\nabla\Lcdot(h\bar{\bm u})$ in view of (2.5).
After some lengthy calculations, we arrive at the following evolution equation for $\bar{\bm u}$:
$$\left(1+{\delta^2\over h}{\cal L}(h, b)\right)\bar{\bm u}_t+\epsilon(\bar{\bm u}\Lcdot\nabla)\bar{\bm u}+\nabla\eta
={\epsilon\delta^2\over 3h}\nabla\Big[h^3\{(\bar{\bm u}\Lcdot\nabla)\nabla\Lcdot\bar{\bm u}-(\nabla\Lcdot\bar{\bm u})^2\}\Big]+\epsilon\delta^2Q, \eqno(3.20a)$$
with
$$Q=-{\beta\over 2h}\Big[\nabla\{h^2\bar{\bm u}\Lcdot\nabla(\nabla b\Lcdot\bar{\bm u})\}-h^2\{\bar{\bm u}\Lcdot\nabla(\nabla\Lcdot\bar{\bm u})-(\nabla\Lcdot\bar{\bm u})^2\}\nabla b\Big]
-\beta^2\{(\bar{\bm u}\Lcdot\nabla)^2b\}\nabla b, \eqno(3.20b)$$
where ${\cal L}(h, b)$ is a linear differential operator defined by
$${\cal L}(h, b){\bm f}=-{1\over 3}\nabla (h^3\nabla\Lcdot{\bm f})+{\beta\over 2}\{\nabla(h^2\nabla b\Lcdot{\bm f})-h^2\nabla b(\nabla\Lcdot{\bm f})\}+\beta^2h\nabla b(\nabla b\Lcdot{\bm f}), \eqno(3.20c)$$
for any vector function ${\bm f}\in \mathbb{R}^2$. This equation coincides perfectly with that obtained by different methods.
See Green \& Naghdi [4], Miles \& Salmon [5], Bazdenkov  {\it et al}. [6] and Lannes \& Bonneton [8].  \par
\medskip
\noindent{(c) Remark}\par
\medskip
\noindent As already demonstrated in \S 2(c), the $\delta^{2n}$ models  with even $n$ have singularities in their linear dispersion relations, although the dispersion characteristics for 
small values of the dispersion parameter have been improved considerably when compared with those of 
the original GN model.  For example, the dispersion relation for the $\delta^4$ model exhibits a singularity at $k\delta \simeq 4.19$, and this feature may limit the range of
applicability of the model.
 The  simplest extended GN model which avoids
this undesirable behavior in higher wavenumber is the 1D $\delta^6$ model with a flat bottom topography. Its derivation can be made straightforwardly by means of the 
procedure developed in this section.  Therefore, without entering into the detailed analysis, we describe only the final result.
 \par
The evolution equation for ${\bar u}$  which extends equation $(3.7b)$ to order $\delta^6$ can now be written in the form 
$$\bar u_t+\epsilon\bar u\bar u_x+\eta_x={\delta^2\over 3h}\left\{h^3(\bar u_{xt}+\epsilon\bar u\bar u_{xx}-\epsilon\bar u_x^2)\right\}_x$$
$$+{\delta^4\over 45h}\left[\left\{h^5(\bar u_{xxt}+\epsilon\bar u\bar u_{xxx}-5\epsilon \bar u_x\bar u_{xx})\right\}_x-3\epsilon h^5\bar u_{xx}^2\right]_x$$
$$+{\delta^6\over 945h}\Bigl[\{h^7(2\bar u_{xxxxt}+2\epsilon\bar u\bar u_{xxxxx}-14\epsilon\bar u_x\bar u_{xxxx}-30\epsilon\bar u_{xx}\bar u_{xxx})\}_x $$
$$+\{h^6h_x(14\bar u_{xxxt}+14\epsilon\bar u\bar u_{xxxx}-112\epsilon\bar u_x\bar u_{xxx}+42\epsilon\bar u_{xx}^2)\}_x$$
$$+\{h^5(hh_x)_x(7\bar u_{xxt}+7\epsilon\bar u\bar u_{xxx}-63\epsilon\bar u_x\bar u_{xx})\}_x
+\epsilon\{10h^7\bar u_{xxx}^2-35h^5(hh_x)_x\bar u_{xx}^2\}\Bigr]_x. \eqno(3.21)$$
The linear dispersion relation for the system of equations $(3.7a)$ and (3.21)  is then given by
$$\omega^2={k^2\over 1+{1\over 3}(k\delta)^2-{1\over 45}(k\delta)^4+{2\over 945}(k\delta)^6}. \eqno(3.22)$$
See (2.51) with $D_6$ from (2.50).
Obviously, the singularity does not occur in $\omega$ for arbitrary values of $k\delta$, as opposed  to the $\delta^4$ model. This ensures the
well-posedness of the system of linearized equations for the  model. 
Another way to improve the linear dispersion characteristics of the $\delta^4$ model is to replace the depth-averaged velocity by the fluid velocity
evaluated at a certain depth. This procedure, however, leads to the violation
 of the conservation of total energy of the $\delta^4$ model. See Madsen $\&$  
Sch\"affer [17], for example. Refer also to Bona {\it et al} [28] as for a mathematical analysis on the linear properties of higher-order shallow-water models.
In any case,  it is an interesting issue to explore various features of the $\delta^6$ model in comparison with  those of the $\delta^4$ model and its modified versions, 
as well as those of the $\delta^2$ (or GN) model. \par
\bigskip
\noindent{\bf 4. Hamiltonian structure}\par
\medskip
\noindent{(a) Hamiltonian} \par
\medskip
\noindent In this section, we show that the 2D extended GN system derived in \S 2 can be formulated as a Hamiltonian form.
 First, recall that the basic Euler system of equations (1.1)-(1.4) conserves the total energy (or Hamiltonian) $H$ which is the sum of the kinetic energy $K$
and the potential energy $U$:
$$H=K+U={\epsilon^2\over 2}\int_{\mathbb{R}^2}\left[\int_{-1+\beta b}^{\epsilon\eta}\left\{(\nabla \phi)^2
+{1\over\delta^2}\phi_z^2\right\}{\mathrm d}z\right]{\mathrm d}{\bm x}+{\epsilon^2\over 2}\int_{\mathbb{R}^2}\eta^2{\mathrm d}{\bm x}. \eqno(4.1)$$
As pointed out by Zakharov [9], the constancy of $H$ can be confirmed by direct computation using (1.1)-(1.4) coupled with the boundary conditions (1.5).  \par
The integrand of $K$ is then modified, after using  (1.1) and  (1.4), as well as the definitions (2.2), (2.3) and (2.10), as  
$$\int_{-1+\beta b}^{\epsilon\eta}\left\{(\nabla \phi)^2+{1\over\delta^2}\phi_z^2\right\}{\mathrm d}z
=\nabla\Lcdot\left[\int_{-1+\beta b}^{\epsilon\eta}\phi\nabla\phi\,{\mathrm d}z\right]-\epsilon\psi\nabla\eta\Lcdot{\bm u}+{1\over\delta^2}\psi w. \eqno(4.2)$$
Substituting $w$ from (2.4) into (4.2), the expression of the kinetic energy $K$ reduces, after performing the integration with respect to ${\bm x}$ under the boundary condition (1.5), to
$$K={\epsilon^2\over 2}\int_{\mathbb{R}^2}[h\bar{\bm u}\Lcdot\nabla\psi]{\mathrm d}{\bm x}. \eqno(4.3)$$
Thus, the Hamiltonian can be rewritten in a simple form
$$H={\epsilon^2\over 2}\int_{\mathbb{R}^2}\left[h\bar{\bm u}\Lcdot\nabla\psi+{1\over\epsilon^2}(h-1+\beta b)^2\right]{\mathrm d}{\bm x}, \eqno(4.4)$$
where we have replaced $\eta$ by $(h-1+\beta b)/\epsilon$ in the expression of the potential energy.
The quantity $\nabla\psi(={\bm V})$ expressed in terms of $h$ and $\bar{\bm u}$ is available in the form of a series expansion. See (2.35) for the expression of ${\bm V}$ up to order $\delta^2$.
Inserting this into (4.4), we obtain a series expansion of $H$ in powers of $\delta^2$
$$H=\epsilon^2\sum_{n=0}^\infty\delta^{2n}H_n, \eqno(4.5a)$$
with the first two of $H_n$ being given by
$$H_0={1\over 2}\int_{\mathbb{R}^2}\left[h{\bar{\bm u}}^2+{1\over\epsilon^2}(h-1+\beta b)^2\right]{\mathrm d}{\bm x}, \eqno(4.5b)$$
$$H_1={1\over 6}\int_{\mathbb{R}^2}\Bigl[h^3(\nabla\Lcdot\bar{\bm u})^2-3\beta h^2(\nabla b\Lcdot\bar{\bm u})\nabla\Lcdot \bar{\bm u}+3\beta^2h(\nabla b\Lcdot\bar{\bm u})^2\Bigr]{\mathrm d}{\bm x}. \eqno(4.5
c)$$
 \par
\medskip
\noindent{(b)  Momentum density} \par
\medskip
\noindent In formulating the extended  GN system as a Hamiltonian form, it is crucial to introduce 
the momentum density ${\bm m}$. It is given by the following relation
$$\epsilon{\bm m}={\delta H\over\delta\bar{\bm u}}, \eqno(4.6)$$
where the operator $\delta/\delta\bar{\bm u}$ is the variational derivative defined by
$${\partial\over\partial\epsilon}H(\bar{\bm u}+\epsilon {\bm w})\big|_{\epsilon=0}=\int_{\mathbb{R}^2}{\delta H\over\delta\bar{\bm u}}\Lcdot{\bm w\,}{\mathrm d}{\bm x}, \eqno(4.7)$$
for arbitrary vector function ${\bm w}\in \mathbb{R}^2$. As seen from (4.5) and its higher-order analog, the integrand of  $K$ is quadratic in $\bar{\bm u}$, and hence $K$ obeys the scaling law
$$K(\epsilon\bar{\bm u}, h, b)=\epsilon^2K(\bar{\bm u}, h, b). \eqno(4.8)$$
Putting ${\bm w}=\bar{\bm u}$ in (4.7) and noting that the potential energy $U$ is independent of $\bar{\bm u}$, we see that
$${\partial\over\partial\epsilon}K((1+\epsilon)\bar{\bm u}, h, b)\big|_{\epsilon=0}=\int_{\mathbb{R}^2}{\delta K\over\delta\bar{\bm u}}\Lcdot\bar{\bm u}\,{\mathrm d}{\bm x}
=\int_{\mathbb{R}^2}{\delta H\over\delta\bar{\bm u}}\Lcdot\bar{\bm u}\,{\mathrm d}{\bm x}. \eqno(4.9)$$
On the other hand, in view of (4.8), ${\partial\over\partial\epsilon}K((1+\epsilon)\bar{\bm u}, h, b))\big|_{\epsilon=0}=2K(\bar{\bm u}, h, b)$. Hence, (4.9) gives, after introducing ${\bm m}$ from (4.6),
$K={\epsilon\over 2}\int_{\mathbb{R}^2}{\bm m}\Lcdot\bar{\bm u}\,{\mathrm d}{\bm x}$, so that $H$ is expressed compactly as
$$H={1\over 2}\int_{\mathbb{R}^2}[\epsilon\,{\bm m}\Lcdot\bar{\bm u}+(h-1+\beta b)^2]\,{\mathrm d}{\bm x}. \eqno(4.10)$$
Comparing (4.4) and (4.10), we obtain  the key relation which connects the variable $\nabla\psi$ with the momentum density ${\bm m}$:
$${\bm m}=\epsilon h\nabla\psi. \eqno(4.11)$$
\par
We can compute the momentum density in accordance with the definition (4.6). For example, the approximate expression of ${\bm m}$  which takes into account  the $\delta^2$ terms follows from (4.5). Explicitly,
$${\bm m}=\epsilon\Bigl[h\bar{\bm u}+\delta^2\Big\{-{1\over 3}\nabla(h^3\nabla\Lcdot \bar{\bm u})+{\beta\over 2}\nabla(h^2\nabla b\Lcdot\bar{\bm u})-{\beta\over 2} h^2(\nabla\Lcdot \bar{\bm u})\nabla b$$
$$+3\beta^2h(\nabla b\Lcdot\bar{\bm u})\nabla b\Big\}\Bigr]+O(\delta^4). \eqno(4.12)$$
Inverting this relation, we obtain the expression of $\bar{\bm u}$ in terms of $h$ and ${\bm m}$ which, substituted into (4.10), yields the Hamiltonian as a functional of $h$ and ${\bm m}$.
Note that the kinetic energy obeys the scaling law $K(\epsilon{\bm m}, h, b)=\epsilon^2K({\bm m}, h, b)$, and hence $K={1\over 2}\int_{\mathbb{R}^2}\delta H/\delta{\bm m}\cdot{\bm m}\,{\mathrm d}{\bm x}$.
Comparing this expression with $K={\epsilon\over 2}\int_{\mathbb{R}^2}{\bm m}\Lcdot\bar{\bm u}\,{\mathrm d}{\bm x}$, we obtain
the dual relation to (4.6)
$$\epsilon\bar{\bm u}={\delta H\over\delta {\bm m}}. \eqno(4.13)$$
\par
\medskip
\noindent{(c)  Evolution equation for the momentum density} \par
\medskip
\noindent To derive the evolution equation for the momentum density ${\bm m}$, we first establish the following formula which provides
 the variational derivative of $H$ with respect to $h$:
$${\delta H\over\delta h}=\epsilon^2\left({1\over 2}{\bm u}^2+{w^2\over 2\delta^2}-{\bm u}\Lcdot\bar{\bm u}+hw\nabla\Lcdot\bar{\bm u}-\beta w\nabla b\Lcdot\bar{\bm u}\right)+h-1+\beta b. \eqno(4.14)$$
A proof of (4.14) is given in Appendix C. \par
Now, we proceed to derive the evolution equation for ${\bm m}$. We start from the evolution equation for ${\bm V}$ from (2.9).  First, we multiply (2.9) by $ h$ and use equation (2.5)
to recast it into the form 
$$(h{\bm V})_t+\epsilon\nabla(h\bar{\bm u}\Lcdot{\bm V})+\epsilon(\nabla\Lcdot h\bar{\bm u}){\bm V}-\epsilon(\bar{\bm u}\Lcdot{\bm V})\nabla h$$
$$+\epsilon h\nabla\left[({\bm u}-\bar{\bm u})\Lcdot{\bm V}-{1\over 2}{\bm u}^2-{1\over 2\delta^2}w^2\right]+h\nabla\eta={\bm 0}. \eqno(4.15)$$
It follows from (2.4), (2.8) and (4.14) that
$$({\bm u}-\bar{\bm u})\Lcdot{\bm V}-{1\over 2}{\bm u}^2-{1\over 2\delta^2}w^2={1\over\epsilon^2}{\delta H\over\delta h}-{1\over\epsilon^2}(h-1+\beta b). \eqno(4.16)$$
Introducing (4.16) into (4.15) and using the relation ${\bm m}=\epsilon h{\bm V}$ from (4.11), we obtain the evolution equation for  ${\bm m}$
$${\bm m}_t+\epsilon\nabla(\bar{\bm u}\Lcdot{\bm m})+\epsilon^2\nabla\Lcdot(h\bar{\bm u}){\bm V}-\epsilon^2(\bar{\bm u}\Lcdot{\bm V})\nabla h+h\nabla\left({\delta H\over\delta h}\right)={\bm 0}. \eqno(4.17)$$
A slight modification of equation (4.17) is possible if one notes the formula $\nabla\Lcdot(h\bar{\bm u})=h\nabla\Lcdot \bar{\bm u}+\nabla h\Lcdot\bar{\bm u}$ and the relation ${\bm V}={\bm m}/(\epsilon h)$.
This leads to
$${\bm m}_t+\epsilon\nabla(\bar{\bm u}\Lcdot{\bm m})+\epsilon(\nabla\Lcdot \bar{\bm u}){\bm m}
+{\epsilon\over h}\{(\nabla h\Lcdot\bar{\bm u}){\bm m}-(\bar{\bm u}\Lcdot{\bm m})\nabla h\} +h\nabla\left({\delta H\over\delta h}\right)={\bf 0}. \eqno(4.18)$$
Furthermore, if we divide (4.18) by $h$ and use (2.5), we can put it in the form of local conservation law
$$\left({{\bm m}\over h}\right)_t+\nabla\left({\epsilon \bar{\bm u}\Lcdot{\bm m}\over h}+{\delta H\over\delta h}\right)={\bm 0}. \eqno(4.19)$$
\par
 As already discussed in a sentence below equation (4.12), the velocity $\bar{\bm u}$ in equation (4.18) can be expressed in terms of $h$ and ${\bm m}$. Thus, the resulting evolution equation, when coupled with equation (2.5)
for $h$, constitutes a closed system of equations for $h$ and ${\bm m}$. This system is equivalent to the extended GN system and will be used for establishing the Hamiltonian formulation of the latter system. \par
\medskip
\noindent{(d)  Hamiltonian formulation} \par
\medskip
\noindent In this section, we demonstrate that the 2D extended GN system can be formulated as a Hamiltonian system. To this end, we  introduce the noncanonical 
 Lie-Poisson bracket between any pair of smooth functional $F$ and $G$
$$\{F, G\}=-\int_{\mathbb{R}^2}\left[\sum_{i, j=1}^2{\delta F\over\delta m_i}(m_j\partial_i+\partial_jm_i){\delta G\over\delta m_j}+h\,{\delta F\over\delta {\bm m}}\Lcdot\nabla{\delta G\over\delta h}
+{\delta F\over\delta h}\,\nabla\Lcdot\left(h\,{\delta G\over\delta {\bm m}}\right)\right]{\mathrm d}{\bm x}, \eqno(4.20)$$
where we have put ${\bm m}=(m_1, m_2)$ and $\partial_1=\partial/\partial x, \partial_2=\partial/\partial y$.
Note that the partial derivatives $\partial_i\ (i=1, 2)$ operate on all terms they multiply to the right. Then, our main result is given by
the following theorem. \par
\bigskip
{\bf Theorem 1.} {\it The 2D extended GN system (2.5) and (2.9) (or equivalently, (4.18)) can be written in the form of Hamilton's equations }
$$h_t=\{h, H\}, \eqno(4.21a)$$
$$ m_{i,t}=\{m_i, H\}, \quad (i=1, 2). \eqno(4.21b)$$ 
\par
\bigskip
\noindent{\it Proof.}  The identification $F=h$ and $G=H$ in (4.20) gives $\{h, H\}=-\nabla\Lcdot\left(h\,{\delta H\over\delta {\bm m}}\right)$. By virtue of (4.13), this becomes $-\epsilon\nabla\Lcdot(h\bar{\bm u})$.
Hence, equation $(4.21a)$ is expressed as $h_t=-\epsilon\nabla\Lcdot(h\bar{\bf u})$, which coincides with equation (2.5). \par
The equation of motion for $m_i$ $(4.21b)$ with (4.20) can be written explicitly as
$$m_{i,t}=-\sum_{j=1}^2(m_j\partial_i+\partial_jm_i){\delta H\over\delta m_j}-h\partial_i{\delta H\over\delta h}$$
$$=-\partial_i\left({\bm m}\Lcdot{\delta H\over\delta {\bm m}}\right)-m_i\nabla\Lcdot{\delta H\over\delta {\bm m}}+\sum_{j=1}^2\left({\partial m_j\over\partial x_i}-{\partial m_i\over\partial x_j}\right){\delta H\over\delta m_j}
-h\partial_i{\delta H\over\delta h}, \quad (i=1, 2). \eqno(4.22)$$
Using the relations $\delta H/\delta{\bm m}=\epsilon\bar{\bm u}$ and ${\bm m}=\epsilon h\nabla\psi=\epsilon h{\bm V}$, we deduce
$$\sum_{j=1}^2\left({\partial m_j\over\partial x_i}-{\partial m_i\over\partial x_j}\right){\delta H\over\delta m_j}
={\epsilon\over h}\left[(\bar{\bm u}\Lcdot{\bm m}){\partial h\over\partial x_i}-(\bar{\bm u}\Lcdot\nabla h)m_i\right]. \eqno(4.23)$$
Substituting (4.23) into (4.22) and rewriting the resultant equation in the vectorial representation, we can see that ${\bm m}$ evolves according to the equation
$${\bm m}_t=-\epsilon\nabla(\bar{\bm u}\Lcdot{\bm m})-\epsilon(\nabla\Lcdot \bar{\bm u}){\bm m}
-{\epsilon\over h}\{(\nabla h\Lcdot\bar{\bm u}){\bm m}-(\bar{\bm u}\Lcdot{\bm m})\nabla h\}- h\nabla\left({\delta H\over\delta h}\right),  \eqno(4.24)$$
which is just equation (4.18). This completes the proof of Theorem 1.  \hspace{\fill}$\blacksquare$ \par
\medskip
We recall that the bracket (4.20) has been introduced by Holm [7] to formulate the 2D GN equations as a Hamiltonian system. 
Combining this fact  with Theorem 1, we conclude that the extended GN system has the same Hamiltonian structure as that of the GN system.
Hence, its truncated version like the $\delta^{2n}$ model shares the same property. 
As for the Hamiltonian formulation of the GN system,  see also Camassa  {\it et al}. [29] for the 2D case, and  Constantin [30]
and Li [31] for the 1D case. \par
\medskip
\noindent{(e) Remark} \par
\medskip
\noindent For the ${\delta^4}$ model discussed in \S 3,  the approximate Hamiltonian is given by (3.12).
It follows from (4.6) that
$${\bm m}=\epsilon\left[h\bar{\bm u}-{\delta^2\over 3}\,\nabla(h^3\nabla\Lcdot\bar{\bm u})-{\delta^4\over 45}\,\nabla[\nabla\Lcdot\{h^5\nabla(\nabla\Lcdot\bar{\bm u})\}\right]+O(\delta^6), \eqno(4.25)$$
and the variational derivative of $H$ with respect to $h$ under fixed ${\bm m}$ is found to be as
$${\delta H\over\delta h}=h-1-\epsilon^2\left[{1\over 2}{\bar{\bm u}}^2+{\delta^2\over 2}h^2(\nabla\Lcdot\bar{\bm u})^2-{\delta^4\over 18}h^4\{\nabla(\nabla\Lcdot\bar{\bm u})\}^2\right]
+O(\delta^6). \eqno(4.26)$$
Note that ${\bm m}$ is  also computed from (2.24) with the aid of  the formula ${\bm m}=\epsilon h{\bm V}$
 whereas the expression of  $\delta H/\delta h$ in terms of $h$ and $\bar{\bm u}$ can be derived  by substituting (2.22) and (2.23) into (4.14) with $b=0$.
Introducing (4.25) and (4.26) into equation $(4.21b)$ and using equation $(4.21a)$ (or (2.5)) to replace the time derivative of $h$ by $-\epsilon\nabla\Lcdot(h\bar{\bm u})$, we obtain the evolution equation for $\bar{\bm u}$.
One can  confirm by a direct computation that it reproduces equation (3.3).  We point out that thanks to (4.19), the integral $\int_{\mathbb{R}^2}{{\bm m}\over h}{\mathrm d}{\bm x}$ 
becomes a constant of motion. When the approximate expression of ${\bm m}$ from (4.25) is used in this integral, it yields
the conserved quantity ${\bm L}$ given by (3.13).
\par
For the 2D GN model with an uneven bottom topography, one needs the Hamiltonian truncated at order $\delta^2$, which is available in (4.5). Introducing this expression into (4.21), the equations of
motion  for $h$ and ${\bm m}$ are found to coincide with the GN system (2.5) and (3.20), as observed for the first time by Holm [7].
\par
In the 1D case, the evolution equation for ${\bm m}$ from $(4.21b)$ (or (4.18)) simplifies to
$$m_t+\epsilon(\bar um)_x+\epsilon \bar u_xm+h\left({\delta H\over \delta h}\right)_x=0. \eqno(4.27)$$
For a flat bottom topography, the approximate Hamiltonian correct up to order $\delta^6$ can be expressed as (Matsuno [1])
$$H=\epsilon^2\int^\infty_{-\infty}\Biggl[{1\over 2}\left\{h\bar u^2+{1\over\epsilon^2}(h-1)^2\right\} +{\delta^2\over 6}h^3\bar u_x^2
-{\delta^4\over 90}h^5\bar u_{xx}^2$$
$$+{\delta^6\over 1890}\left\{2h^7\bar u_{xxx}^2-7h^5(hh_x)_x\bar u_{xx}^2\right\}+O(\delta^8)\Biggr]{\mathrm d}x. \eqno(4.28)$$
The momentum density $m$ computed from the 1D version of (4.6) with $H$ from (4.28) is
$$m=\epsilon\left\{h\bar u-{\delta^2\over 3}(h^3\bar u_x)_x-{\delta^4\over 45}(h^5\bar u_{xx})_{xx}\right\}
-{\epsilon\delta^6\over 945}\Bigl[7\{h^5(hh_x)_x\bar u_{xx}\}_{xx}+2(h^7\bar u_{xxx})_{xxx}\Bigr]+O(\delta^8), \eqno(4.29)$$
and the variational derivative $\delta H/\delta h$ with $m$ being fixed is found to be
$${\delta H\over \delta h}=h-1-{\epsilon^2\over 2}\left(\bar u^2+\delta^2h^2\bar u_x^2-{\delta^4\over 9}h^4\bar u_{xx}^2\right)$$
$$+{\epsilon^2\delta^6\over 270}\left\{(10h^5h_{xx}+25h^4h_x^2)\bar u_{xx}^2
+10h^5h_x(\bar u_{xx}^2)_x+2h^6\bar u_{xx}\bar u_{xxxx}\right\}+O(\delta^8). \eqno(4.30)$$
Substituting (4.29) and (4.30) into equation (4.27) and using equation $(4.21a)$ (or $(3.7a)$) to replace $h_t$ by $-\epsilon(h\bar u)_x$,  we see that equation (4.27) reproduces   equation (3.21). \par
\bigskip
\noindent{\bf 5. Relation to Zakharov's Hamiltonian formulation}\par
\medskip
\noindent{(a) Zakharov's formulation } \par
\noindent Zakharov [9]  (see also Zakharov \& Kuznetsov [32]) showed that the water wave problem (1.1)-(1.5) permits a canonical Hamiltonian formulation
in termas of the canonical variables $\eta$ and $\psi$. 
If we change the variables from $(\eta, \psi)$ to $(h, \nabla\psi)$,
then  the equations of motion for the  variables $h$ and $\nabla\psi$
are written in the form
$$h_t=-{1\over\epsilon}\nabla\Lcdot{\delta H\over \delta \nabla\psi}, \quad
 \nabla\psi_t=-{1\over\epsilon}\nabla{\delta H\over \delta h}, \eqno(5.1)$$
 where the Hamiltonian $H$ is given by (4.1).
 If we define the  Poisson bracket between any pair of smooth functionals $F$ and $G$  by
 $$\{F, G\}=-{1\over\epsilon}\int_{\mathbb{R}^2}\left[{\delta F\over\delta h}\left(\nabla\Lcdot{\delta G\over\delta\nabla\psi}\right)
-\left(\nabla\Lcdot{\delta F\over\delta\nabla\psi}\right){\delta G\over\delta h}\right]{\mathrm d}{\bm x}, \eqno(5.2)$$
then the system of equations (5.1) can be put into the form of Hamilton's equations
$$h_t=\{h, H\}, \quad \nabla\psi_t=\{\nabla\psi, H\}. \eqno(5.3)$$
\bigskip
\noindent{(b) Transformation of the Zakharov system to the extended GN system} \par
\medskip
\noindent Here, we establish the following theorem. \par
\medskip
{\bf Theorem 2.} {\it Zakharov's system of equations (5.3)  is equivalent to the extended GN system (4.21).}\par
\bigskip
\noindent{\it Proof.} We change  the variable $\nabla\psi$ to the momentum density ${\bm m}$ whereas  $h$ remains the common variable for both systems,
and take the variation of any smooth functional $F$ in two alternative ways. This gives
$$\int_{\mathbb{R}^2}\left[{\delta F\over\delta h}\bigg|_{\nabla\psi}\delta h+{\delta F\over\delta \nabla\psi}\bigg|_h\Lcdot\delta\nabla\psi\right]{\mathrm d}{\bm x}
=\int_{\mathbb{R}^2}\left[{\delta F\over\delta h}\bigg|_{\bm m}\delta h+{\delta F\over\delta {\bm m}}\bigg|_h\Lcdot\delta{\bm m}\right]{\mathrm d}{\bm x}. \eqno(5.4)$$
Using (4.11), the variation of ${\bm m}$ is found to be as $\delta{\bm m}=\epsilon\nabla\psi\delta h+\epsilon h\delta\nabla\psi$.
 We substitute this expression into the right-hand side of (5.4)
and then compare the coefficients  of $\delta h$ and $\delta\nabla\psi$ on both sides. It follows from the coefficient of $\delta h$ that 
$${\delta F\over\delta h}\bigg|_{\nabla\psi}={\delta F\over\delta h}\bigg|_{\bm m}+\epsilon{\delta F\over\delta {\bm m}}\bigg|_h\Lcdot\nabla\psi
={\delta F\over\delta h}\bigg|_{\bm m}+{1\over h}{\delta F\over\delta {\bm m}}\bigg|_h\Lcdot{\bm m}, \eqno(5.5)$$
where in the last line, we have used (4.11) again, and the coefficient of $\delta\nabla\psi$ yields
$${\delta F\over\delta \nabla\psi}\bigg|_h=\epsilon h{\delta F\over\delta {\bm m}}\bigg|_h. \eqno(5.6)$$
If we put $F=H$ in (5.5) and (5.6) and use (4.11), we obtain the relations
$${\delta H\over\delta h}\bigg|_{\nabla\psi}={\delta H\over\delta h}\bigg|_{\bm m}+{\epsilon\bar{\bm u}\Lcdot{\bm m}\over h}, \quad
{\delta H\over\delta \nabla\psi}\bigg|_h=\epsilon^2h\bar{\bm u}. \eqno(5.7)$$
\par
To proceed, we introduce the second relation in (5.7)  into the first equation in   (5.1) and see that it coincides with equation (2.5), or equivalently equation $(4.21a)$. 
On the other hand, substitution of $\nabla\psi$ from (4.11)  and the first relation in (5.7) into the second equation in  (5.1) recasts it into the form
$$\left({\bm m}\over h\right)_t=-\nabla\left({\delta H\over\delta h}\bigg|_{\bm m}\right)-\epsilon\nabla\left({\bar{\bm u}\Lcdot{\bm m}\over h}\right), \eqno(5.8)$$
which is nothing but equation (4.19), or equivalently equation $(4.21b)$. 
Thus, we have established the equivalence of the Zakharov system and the extended GN system. \par
 Last, we  show that the  bracket (5.2) transforms to the   bracket (4.20)
under the change of  variables $(h, \nabla\psi) \rightarrow (h, {\bm m})$. 
 To this end, we substitute (5.5) and (5.6) into (5.2) and integrate the second term in the brackets by parts to transform it into the form
 $$\{F, G\}=-\int_{\mathbb{R}^2}\biggl[{\delta F\over\delta h}\nabla\Lcdot\left(h{\delta G\over\delta {\bm m}}\right)
+\left({\bm m}\Lcdot{\delta F\over\delta {\bm m}}\right)\left(\nabla\Lcdot{\delta G\over\delta {\bm m}}\right)
+{1\over h}\left({\bm m}\Lcdot{\delta F\over\delta {\bm m}}\right)\left(\nabla h\Lcdot{\delta G\over\delta {\bm m}}\right)$$
$$+h{\delta F\over\delta {\bm m}}\Lcdot\nabla{\delta G\over\delta h}+{\delta F\over\delta {\bm m}}\Lcdot\nabla\left({\bm m}\Lcdot{\delta G\over\delta {\bm m}}\right)
-{1\over h}\left(\nabla h\Lcdot{\delta F\over\delta {\bm m}}\right)\left({\bm m}\Lcdot{\delta G\over\delta {\bm m}}\right)
\biggr]{\mathrm d}{\bm x}. \eqno(5.9)$$
Now, a straightforward  computation shows that
$$\int_{\mathbb{R}^2}\sum_{i, j=1}^2{\delta F\over\delta m_i}(m_j\partial_i+\partial_jm_i){\delta G\over\delta m_j}\,{\mathrm d}{\bm x}$$
$$=\int_{\mathbb{R}^2}\left[{\delta F\over\delta {\bm m}}\Lcdot\nabla\left({\bm m}\Lcdot{\delta G\over\delta {\bm m}}\right)
+\left({\bm m}\Lcdot{\delta F\over\delta {\bm m}}\right)\left(\nabla \Lcdot{\delta G\over\delta {\bm m}}\right)+
\sum_{i, j=1}^2{\delta F\over\delta m_i}\left({\partial m_i\over\partial x_j}-{\partial m_j\over\partial x_i}\right){\delta G\over\delta m_j}\right]{\mathrm d}{\bm x}. \eqno(5.10)$$
Using the relation $m_i=\epsilon h\partial\psi/\partial x_i$ from (4.11), we can modify the third term in the integrand on the right-hand side of (5.10) as
$$\sum_{i, j=1}^2{\delta F\over\delta m_i}\left({\partial m_i\over\partial x_j}-{\partial m_j\over\partial x_i}\right){\delta G\over\delta m_j}
={1\over h}\left({\bm m}\Lcdot{\delta F\over\delta {\bm m}}\right)\left(\nabla h\Lcdot{\delta G\over\delta {\bm m}}\right)
-{1\over h}\left(\nabla h\Lcdot{\delta F\over\delta {\bm m}}\right)\left({\bm m}\Lcdot{\delta G\over\delta {\bm m}}\right). \eqno(5.11)$$
If we introduce (5.10) and (5.11) into (5.9), then we find that the bracket (5.2) transforms  to the bracket (4.20). This completes the proof of Theorem 2.
 \hspace{\fill}$\blacksquare$  \par
\medskip
\noindent{(c) {Remark} \par
\medskip
\noindent The Lie-Poisson bracket defined by (4.20)  has a skew-symmetry $\{F,G\}=-\{G,F\}$,  
and satisfies the Jacobi identity $\{F,\{G,H\}\}+\{G,\{H,F\}\}+\{H,\{F,G\}\}=0$ 
for any smooth functionals $F, G$ and $H$.  The skew-symmetric nature follows simply by means of the integration by parts.
The Lie-Poisson bracket (4.20) stems naturally
from the canonical Poisson bracket through a sequence of  variable transformations $(\eta, \psi) \rightarrow (h, \nabla\psi) \rightarrow (h, {\bf m})$.
The Jacobi identity can be proved if one applies the same transformations to the Jacobi identity for the canonical Poisson bracket.
 \par
\bigskip
\noindent{\bf 6. Conclusion}\par
\medskip
\noindent In this paper, we have developed a systematic procedure for extending the 2D GN model to include higher-order dispersive effects while
preserving full nonlinearity of the original GN model, and presented various model equations for both  flat and uneven bottom
topographies.  We have also shown that these models permit Hamiltonian formulation. When compared with a previous work (Matsuno [1]), 
the novelties of the present study are summarized as follows:\ (i) The general formulation of the water wave problem in the 3D setting which takes into account
 the effect of uneven bottom topography, (ii) The analysis of the linear dispersion relation for the extended GN system which features
 the linearized GN system, (iii) The derivation of the 1D $\delta^6$ model which is the second example of the GN models with  nonsingular linear dispersion
relation, the first one being the original GN model, (iv) Computation of the pressure distribution at the fluid bottom in the framework of the 2D $\delta^4$ model, 
(v) A simple derivation of the key relation between the momentum density and  gradient of the surface potential which plays the central role in establishing the
Hamiltonian structure of the GN system. \par
There are a number of  interesting problems associated with the extended GN equations that are worthy of further study. In conclusion, we list some of them:  \par
\noindent (i) The identification of physically relevant models among various extended GN equations, \par
\noindent (ii) The effect of higher-order dispersion on the wave characteristics in comparison with that predicted by other asymptotic models like Boussinesq equations, \par
\noindent (iii) Numerical computations of the initial value problems as well as solitary and periodic  wave solutions, \par
\noindent (iv) The justification of the asymptotic models by means of the rigorous mathematical analysis.
\par
\bigskip
\noindent{\bf Appendix A Positivity of $D_{2n}$ with odd $n$}\par
\medskip
\noindent We show that the function $D_{2n}(\kappa)$ defined by (2.49) is positive definite for odd $n$.  More precisely, it is bounded from below by 1. 
To prove this statement,  we begin with the expression of $D_{2n}$ with $n$ replaced by $2n+1$:
$$D_{2(2n+1)}=1+\sum_{r=0}^n{2^{2(2r+1)}\over \{2(2r+1)\}!}\,B_{2r+1}\kappa^{2(2r+1)}-\sum_{r=1}^n{2^{4r}\over (4r)!}\,B_{2r}\kappa^{4r}. \eqno(A\ 1)$$
The following inequality has been established for the Bernoulli numbers $B_r$ (Abramowitz \& Stegun [23])
$${2(2r)!\over (2\pi)^{2r}}{1\over 1-2^{-2r} }< B_r < {2(2r)!\over (2\pi)^{2r}}{1\over 1-2^{-2r+1}},\quad r\geqslant 1. \eqno(A\ 2)$$
Using the lower bound of the inequality $(A\ 2)$ for the second term of $(A\ 1)$ and the upper bound for the third term, respectively, we obtain the estimate
$$D_{2(2n+1)} \geqslant 1+2\sum_{r=0}^n{\lambda^{2r+1}\over 1-2^{-2(2r+1)}}-2\sum_{r=1}^n{\lambda^{2r}\over 1-2^{-4r+1}}, \eqno(A\ 3)$$
where $\lambda=(\kappa/\pi)^2$. If we introduce the function $g=g(\lambda)$ by
$$g=\sum_{r=0}^n{\lambda^{2r}\over 1-2^{-2(2r+1)}}-\sum_{r=1}^n{\lambda^{2r-1}\over 1-2^{-4r+1}}, \eqno(A\ 4)$$
then  $(A\ 3)$ is expressed as $D_{2(2n+1)}\geqslant 1+2\lambda g(\lambda)$. The inequalty $D_{2(2n+1)}\geqslant 1$ follows provided  $g(\lambda)\geqslant 0$ for $\lambda\geqslant 0$, which we shall now demonstrate. \par
For $\lambda$ in the interval $0\leqslant \lambda \leqslant 1$, we multiply the first term of $g$ by $\lambda$ to obtain the inequality
$$g\geqslant \sum_{r=0}^n{\lambda^{2r+1}\over 1-2^{-2(2r+1)}}-\sum_{r=1}^n{\lambda^{2r-1}\over 1-2^{-4r+1}}. \eqno(A\ 5)$$
Adding the term $r=n+1$ in the second sum  of $(A\ 5)$ and replacing the summation index  $r$ by $r+1$, we can derive the inequality
$$g\geqslant \sum_{r=0}^n\left({1\over 1-2^{-4r-2}}-{1\over 1-2^{-4r-3}}\right)\,\lambda^{2r+1}=\sum_{r=0}^n{2^{-4r-3}\over (1-2^{-4r-2})(1-2^{-4r-3})}\,\lambda^{2r+1}\geqslant 0. \eqno(A\ 6)$$
Hence, the assertion $g(\lambda)\geqslant 0$ is true for $0\leqslant \lambda \leqslant 1$.
\par
For  $\lambda>1$, we modify $g$ from $(A\ 4)$ in the form
$$g={4\over 3}+\sum_{r=1}^n{\lambda^{2r-1}\over 1-2^{-4r-2}}\left(\lambda-{1-2^{-4r-2}\over 1-2^{-4r+1}}\right). \eqno(A\ 7)$$
However, because
$${1-2^{-4r-2}\over 1-2^{-4r+1}}=1+{7\over 4}\,{2^{-4r}\over 1-2^{-4r+1}}\leqslant {9\over 8}, \quad r\geqslant 1,\eqno(A\ 8)$$
we see that $g(\lambda)>0$ for $\lambda>9/8$. For $\lambda$ in the interval $1<\lambda\leqslant 9/8$, we put $\lambda=1$ in parentheses of $(A\ 7)$
and take into account $(A\ 8)$ to show that $g$ is bounded from below by  the inequality
$$g>{4\over 3}-{7\over 4}\,\sum_{r=1}^n{2^{-4r}\lambda^{2r-1}\over (1-2^{-4r-2})(1-2^{-4r+1})}. \eqno(A\ 9)$$
If we notice that the inequality $(1-2^{-4r-2})(1-2^{-4r+1})\geqslant (1-2^{-6})(1-2^{-3})$ holds  for $r\geqslant 1$, we can simplify $(A\ 9)$ as
$$g>{4\over 3}-{7\over 4}\,{64\over 63}\,{8\over 7}\,{1\over \lambda}\sum_{r=1}^n\left({\lambda\over 4}\right)^{2r}. \eqno(A\ 10)$$
Last, we take the limit $n\rightarrow \infty$ and then put $\lambda=9/8$ in $(A\ 10)$ to obtain a lower bound of $g$:
$$g>{4\over 3}-{128\over 63}{1\over \lambda}{\left({\lambda\over 4}\right)^2\over 1-\left({\lambda\over 4}\right)^2}\geqslant {4\over 3}-{1024\over 6601}\simeq
 1.178, \eqno(A\ 11)$$
indicating that $g(\lambda)>0$ for $1<\lambda\leqslant 9/8$. When combined with the previous result, we see that the inequality $g(\lambda)>0$ holds for $\lambda>1$,
as well as for $0\leqslant \lambda \leqslant 1$.
This completes the proof of the positivity of $D_{2n}$ with odd $n$. 
It is important that the present proof establishes the more stringent result $D_{2(2n+1)}(k\delta)\geqslant 1$. Using this inequality in (2.51), we obtain the estimate 
 $\omega/k=\sqrt{1/D_{2(2n+1)}(k\delta)}\leqslant 1$ for $k\delta\geqslant 0$. 
\par
\bigskip
\noindent{\bf Appendix B  Derivation of $(3.14)$}
\par
\medskip
In accordance with Bernoulli's law, the dimensionless pressure $P$ in the fluid which is scaled by the quantity $\rho gh_0$\ ($\rho$: constant fluid density) is given by
$$P({\bm x}, z, t)=-z-\epsilon\left[\phi_t+{\epsilon\over 2}\left\{(\nabla \phi)^2+{1\over\delta^2}\phi_z^2\right\}\right]. \eqno(B\ 1)$$
At the fluid bottom $z=-1$, $\phi_z=0$  from (1.4), and $\phi_t=f_t, \nabla \phi=\nabla f$ by (2.16). Inserting these relations into $(B\ 1)$, we find that
 the bottom pressure $P_b=P({\bm x}, -1, t)$ has a simple expression in terms of $f$:
$$P_b=1-\epsilon\left[f_t+{\epsilon\over 2}(\nabla f)^2\right]. \eqno(B\ 2)$$
To evaluate the above quantity, we substitute ${\bm V}$ from (2.12) into equation (2.9) and integrate it under the boundary conditions
${\bm u}\rightarrow {\bm 0}, {\bm V}\rightarrow {\bm 0}, w\rightarrow 0, \eta\rightarrow 0, \psi\rightarrow {\rm const.}$ as $|{\bm x}|\rightarrow\infty$.  This gives the evolution equation for $\psi$
$$\psi_t+\epsilon \left({\bm u}\Lcdot{\bm V}-{1\over 2}{\bm u}^2-{1\over 2\delta^2}\,w^2+{\eta\over\epsilon}\right)=0. \eqno(B\ 3)$$
The expressions of ${\bm u}, w$ and ${\bm V}$ in terms of $f$ are already given by (2.22), (2.23) and (2.24), respectively, whereas the  expression of $\psi$
follows from (2.10) and (2.16). 
Substituting these expressions into $(B\ 3)$, we obtain  the following formula correct up to order $\delta^4$:
$$f_t+{1\over 2}\epsilon(\nabla f)^2=-\eta+{\delta^2\over 2}\Big[(h^2r)_t+\epsilon\{\nabla f\Lcdot\nabla(h^2r)+(hr)^2\}\Big]$$
$$-{\delta^4\over 24}\Big[(h^4\nabla^2r)_t+\epsilon\nabla f\Lcdot\nabla(h^4\nabla^2r)+6\epsilon h^2\nabla r\Lcdot\nabla(h^2r)
-3\epsilon h^4(\nabla r)^2+4\epsilon h^4r\nabla^2r\Big]+O(\delta^6), \eqno(B\ 4)$$
where $h=1+\epsilon\eta$ and we have put $r=\nabla^2f$ for simplicity. 
The approximate expression of $\nabla f$ in terms of $h$ and $\bar{\bm u}$ which is correct up to order $\delta^2$ can be derived from (2.19) and $(2.20a)$. It reads
$$\nabla f=\bar{\bm u}+{\delta^2h^2\over 6}\nabla(\nabla\Lcdot\bar{\bm u})+O(\delta^4). \eqno(B\ 5)$$
The last step for obtaining the bottom pressure is to introduce $(B\ 5)$ into $(B\ 4)$ whereby we replace $h_t$ by  $-\epsilon\nabla\Lcdot(h\bar{\bm u})$  in view of equation (2.5) and neglect the $\delta^6$ terms.
After some straightforward calculations, we obtain a lengthy expression of the right-hand  side of $(B\ 4)$ in terms of $h$ and $\bar{\bm u}$ which, substituted into $(B\ 2)$,
yields the formula for $P_b$. \par
\bigskip
\noindent{\bf Appendix C  Proof of $(4.14)$}\par
\par
\medskip
 It now follows by taking the variational derivative of $H$ from (4.1) with respect to $h$ that
$${\delta H\over\delta h}={\epsilon^2\over 2}\left[\left\{(\nabla\phi)^2+{1\over\delta^2}\phi_z^2\right\}\bigg|_{z=\epsilon\eta}+{2\over\epsilon^2}(h-1+\beta b)\right]$$
$$+\epsilon^2\int_{\mathbb{R}^2}\left[\int_{-1+\beta b}^{\epsilon\eta}\left\{\nabla\phi\Lcdot\nabla\left({\delta \phi\over\delta h}\right)
+{1\over\delta^2}\phi_z\left({\delta \phi\over\delta h}\right)_z\right\}{\mathrm d}z\right]{\mathrm d}{\bm x}. \eqno(C\ 1)$$
The integrand of the second term of $(C\ 1)$ can be modified by using (1.1) and  integrating by parts under the boundary conditions (1.4) and (1.5). This gives
$${\delta H\over\delta h}={\epsilon^2\over 2}\left\{{\bm u}^2+{1\over\delta^2}w^2+{2\over\epsilon^2}(h-1+\beta b)\right\}
+\epsilon^2\int_{\mathbb{R}^2}\left(-\epsilon\nabla \eta\Lcdot{\bm u}+{w\over\delta^2}\right){\delta \phi\over\delta h}\bigg|_{z=\epsilon\eta}{\mathrm d}{\bm x}, \eqno(C\ 2)$$
where we have used the definition of ${\bm u}$ and $w$ from (2.2) and (2.3), respectively.
The relation below follows from (2.2) and (2.10) coupled with the formula $\delta h({\bm x}, t)/\delta h({\bm x}^\prime, t)=\delta({\bm x}-{\bm x}^\prime)$, where
$\delta({\bm x}-{\bm x}^\prime)$ is the 2D  delta function:
$${\delta \phi({\bm x}, z, t)\over\delta h({\bm x}^\prime, t)}\bigg|_{z=\epsilon\eta}={\delta\psi({\bm x}, t)\over \delta h({\bm x}^\prime, t)}-w\delta({\bm x}-{\bm x}^\prime). \eqno(C\ 3)$$
We substitute $w$ from (2.4) and $(C\ 3)$ into the second term of $(C\ 2)$ and then perform the integration by parts with respect to ${\bm x}$.  By virtue of the formula
$${\delta \nabla\psi ({\bm x}, t)\over \delta h({\bm x}^\prime, t)}=-{{\bm m}\over\epsilon h^2}\,\delta({\bm x}-{\bm x}^\prime)=-{1\over h}({\bm u}+\epsilon w\nabla \eta)\delta({\bm x}-{\bm x}^\prime), \eqno(C\ 4)$$
which comes from (2.11) and (4.11),   we deduce
$$\int_{\mathbb{R}^2}\left(-\epsilon\nabla \eta\Lcdot{\bm u}+{w\over\delta^2}\right){\delta \phi\over\delta h}\bigg|_{z=\epsilon\eta}{\mathrm d}{\bm x}
=-\int_{\mathbb{R}^2}\nabla\Lcdot(h\bar{\bm u}){\delta \phi\over\delta h}\bigg|_{z=\epsilon\eta}{\mathrm d}{\bm x}$$
$$=-\bar{\bm u}\Lcdot({\bm u}+\epsilon w\nabla \eta)+w\nabla\Lcdot(h\bar{\bm u})=-{\bm u}\Lcdot\bar{\bm u}+hw\nabla\Lcdot\bar{\bm u}-\beta w\nabla b\Lcdot\bar{\bm u}, \eqno(C\ 5)$$
which, introduced in $(C\ 2)$, establishes (4.14). \par

\newpage
\leftline{\bf References} \par
\baselineskip=4mm
\begin{enumerate}[{1.}]
\item  Matsuno, Y. 2015 Hamiltonian formulation of the extended Green-Naghdi equations. {\it Physica D} {\bf 301-302},  1-7.
\item   Serre, F. 1953 Contribution \`a l'\'{e}tude des \'{e}coulements permanents et variables dans les canaux.   {\it Houille Blanche} {\bf 8}, 374-388.
\item   Su, C.H. \&  Gardner, C.S. 1969  Korteweg-de Vries equation and generalizations. III. Derivation of the Korteweg-de Vries equation and Burgers equation.
  {\it J. Math. Phys.}   {\bf 10}, 536-539. 
 \item  Green, A.E. \&  Naghdi, P.M. 1976   A derivation of equations for wave propagation in water of variable depth.  {\it J. Fluid Mech.} {\bf 78},  237-246. 
  \item  Miles, J. \&  Salmon, R. 1985 Weakly dispersive nonlinear gravity waves  {\it J. Fluid Mech.} {\bf 157}, 519-531.
  \item   Bazdenkov, S.V.,  Morozov, N.N. \&  Pogutse, O.P. 1987  Dispersive effects in two-dimensional hydrodynamics. {\it Sov. Phys. Dokl.}  {\bf 32}, 262-264.
  \item   Holm, D.D.  1988 Hamiltonian structure for two-dimensional hydrodynamics with nonlinear dispersion. {\it Phys. Fluids}  {\bf 31}, 2371-2373.
  \item  Lannes, D. \&  Bonneton, P. 2009 Derivation of asymptotic two-dimensional time-dependent equations for surface water wave propagation. {\it Phys. Fluids} {\bf 21}, 016601.
  \item   Zakharov, V.E. 1968 Stability of periodic waves of finite amplitude on the surface of a deep fluid. {\it J. Appl. Mech. Tech. Phys}. {\bf  9}, 190-194.
  \item  Craig, W. \&  Sulem, C. 1993 Numerical simulation of gravity waves. {\it J. Comp. Phys.} {\bf 108}, 73-83. 
  \item   Lannes, D. 2013   {\it Water Waves Problem: Mathematical Analysis and Asymptotics},  Mathematical Surveys and Monographs, vol. 188. American Mathematical Society.
  \item  Whitham, G.B. 1974  {\it Linear and Nonlinear Waves}. New York, NY: John Wiley \& Sons. 
  \item Johnson, R.S. 2002 Camassa-Holm, Korteweg-de Vries and related models for water waves.  {\it J. Fluid Mech.} {\bf 455}, 63-82.
  \item  Constantin, A. \&   Lannes, D. 2009  The hydrodynamic relevance of the Camassa-Holm and Degasperis-Procesi equations.  {\it  Arch. Rat. Mech. Anal.}  {\bf 192}, 165-186.
  \item   Kirby, J.T.  1999  Nonlinear dispersive long waves in water of variable depth.
In  {\it Advances in Fluid Mechanics} {\bf 10} (ed. J.N. Hunt), pp.  55-125, Computational Mechanics Publ.
\item   Madsen, P.A. \& Sch\"affer, H.A.  1998 Higher-order Boussinesq-type equations for surface gravity waves: derivation and analysis.
 {\it Phil. Trans. R. Soc. Lond. A}  {\bf  356}, 3123-3184.
\item   Madsen, P.A. \&  Sch\"affer, H.A. 1999  {\it A review of Boussinesq-type equations for surface gravity waves},
  In  {\it Advances in Coastal and Ocean Engineering 5} (ed. P. Liu), pp. 1-94,   World Scientific.
\item  Witting, J.M.  1984 A unified model for the evolution  of nonlinear water waves. {\it J. Comp. Phys.} {\bf 56}, 203-236.
\item  Dommermuth, D.G. \&  Yue, D.K.P. 1987 A higher-order spectral method for the study of nonlinear gravity waves. {\it J. Fluid Mech.} {\bf 184},  267-288.
\item Castro, A. \& Lannes, D. 2015 Well-posedness and shallow-water stability for a new Hamiltonian formulation of the water waves equations with vorticity.
{\it Indiana Univ. J. of Math.} {\bf 64}, 1169-1270.
\item Gavrilyuk, S., Kalisch, H. \& Khorsand, Z. 2015 A kinetic conservation laws in free surface flow. {\it Nonlinearity} {\bf 28}, 1805-1821. 
\item  Yoon, S.B. \& Liu, P.L.F. 1994 A note on Hamiltonian for long water waves in varying depth. {\it Wave Motion} {\bf 20}, 359-370.
\item  Abramowitz, M. \&  Stegun I.A. (eds.)  1970 Handbook of Mathematical Functions with Formulas, Graphs and Mathematical Tables.  Dover Publications.
\item  Constantin, A.,  Escher, J. \&  Hsu, H.C. 2011 Pressure beneath a solitary water wave: Mathematical theory and experiments. {\it Arch. Rat. Mech. Anal.} {\bf 201}, 251-269.
\item Deconinck, B.,   Oliveras, K.L. \&  Vasan, V.  2012 Relating the bottom pressure and the surface elevation in the water wave problem.  {\it J. Nonl. Math. Phys.} {\bf 19}, Suppl. 1, 1240014.
\item Touboul, J. \& Pelinovsky, E. 2014 Bottom pressure distribution under a solitonic wave reflecting on a vertical wall. {\it Eur. J. Mech. B/Fluids} {\bf 48}, 13-18.
\item  Pelinovsky, E.N.,   Kuznetsov, K.I., Touboul, J. \&  Kurkin, A.A. 2015 Bottom pressure caused by passage of a solitary wave within the strongly nonlinear Green-Naghdi model. {\it Dokl. Phys.} {\bf 60}, 171-174.
\item Bona, J.L., Chen M. \& Saut, J.-C. 2002 Boussinesq equations and other systems for small-amplitude long waves in nonlinear dispersive  media. I: Derivation and linear theory. {\it J. Nonl. Sci.} {\bf 12}, 283-318.
\item   Camassa, R.,  Holm, D.D. \&  Levermore, C.D. 1996 Long-time effects of bottom topography in shallow water.  {\it Physica D} {\bf 98}, 258-286.
\item  Constantin, A.  1997 The Hamiltonian structure of the Camassa-Holm equation. {\it Exp. Math.} {\bf 15}, 53-85.
\item  Li, Y.A. 2002 Hamiltonian structure and linear stability of solitary waves of the Green-Naghdi equations. {\it J. Nonl. Math. Phys.} {\bf 9}, 99-105.
\item  Zakharov, V.E. \& Kuznetsov, E.A. 1997 Hamiltonian formalism for nonlinear waves. {\it Physics-Uspekhi} {\bf 40},  1087-1116.

\end{enumerate}
\end{document}